\documentclass{article}
\usepackage{paper}
\usepackage{mathptmx}
\usepackage[scaled=0.9]{helvet}

\usepackage{graphicx}
\usepackage{caption}

\usepackage[x11names]{xcolor}  
\usepackage{tikz}
\usepackage{fontawesome5}  

\usepackage{subcaption}
\usepackage{verbatim}

\usepackage{bm}  
\usepackage{booktabs}  

\usepackage{tabularx}  
\usepackage{hyperref}  

\usepackage[frozencache, cachedir=_minted]{minted}
\setminted{fontsize=\footnotesize, frame=lines, framesep=2mm}
\usepackage{bytefield}

\usetikzlibrary {positioning,shapes.geometric,shapes.misc,arrows.meta,quotes,math,matrix,fit, backgrounds}

\tikzstyle{every picture}+=[font=\sffamily]

\newcommand{\hquad}{\hspace{0.5em}}

\begin{document}
\title{\texttt{afspm}: A Framework for Manufacturer-Agnostic Automation in Scanning Probe Microscopy}
\newcommand{\myabstract}{Scanning probe microscopy (SPM) is a valuable technique by which one can investigate the physical characteristics of the surfaces of materials.
However, its widespread use is hampered by the time-consuming nature of running an experiment and the significant domain knowledge required.
Recent studies have shown the value of multiple forms of automation in improving this, but their use is limited due to the difficulty of integrating them with SPMs other than the one it was developed for.
With this in mind, we propose an automation framework for SPMs aimed toward facilitating code sharing and reusability of developed components.
Our framework defines generic control and data structure schemas which are passed among independent software processes (components), with the final SPM commands sent after passing through an SPM-specific translator.
This approach permits multi-language support and allows for experimental components to be decoupled among multiple computers.
Our mediation logic limits access to the SPM to a single component at a time, with a simple override mechanism in order to correct detected experiment problems.
To validate our proposal, we integrated and tested it with two SPMs from separate manufacturers, and ran an experiment involving a thermal drift correction component.
}


\newcommand{\hochkomma}{ $^{,}$}  
\author{Nicholas J. Sullivan\thanks{Division of Materials Engineering, Faculty of Engineering, McGill University, Montreal H3A 2T8, Canada}\hochkomma\footnotemark[3], Julio J. Vald\'es\thanks{National Research Council Canada, Digital Technologies Research Centre, Ottawa ON K4A 0S2, Canada}, Kirk H. Bevan\footnotemark[1]\hochkomma\footnotemark[3], and Peter Grutter\thanks{Department of Physics, McGill University, Montreal H3A 2T8, Canada}}
\date{}  
\begin{titlepage}
  \maketitle
  \myabstract
\end{titlepage}

\section{Introduction}
In Scanning Probe Microscopy (SPM), an atomically-sharp tip is scanned above a surface of interest, while measuring one or more properties.
This process allows atomic-level imaging of properties, spectroscopic analysis, and even manipulation of a sample (toward atomic-scale manufacturing) \citep{voigtlander2015scanning}.
However, a number of factors limit wider adoption of these techniques.
Firstly, the preparing, running, and analyzing of such experiments requires significant domain knowledge and expertise, as the SPM systems are highly complicated and prone to various sources of variability .
Secondly, running an individual experiment traditionally involves constant user attention, requiring frequent monitoring and manipulation of SPM parameters.
Lastly, limiting the decision-making of an experiment to the individual choices of a single researcher constrains the statistical understanding of what is being analyzed.

Previous studies have shown the value of automation for improving upon these.
Automatic conditioning of the probe tip has been investigated to ensure proper surface characterization and/or manipulation \citep{rashidi_2018_asm_in-situ_tip_coniditioning_through_ml, krull_2020_deepspm,gordon20_embed_human_heuris_machin_learn}.
Researchers have also automatically classified the structure of a surface, with the aim of automating the next scan or manipulation location.
Generally, this has involved detecting atoms \citep{roccapriore21_reveal_chemic_bondin_adatom_array}, molecules \citep{sotres21_enabl_auton_scann_probe_micros}, or defects of interest \citep{maksov_2019_DL_analysis_of_defect_and_phase_evolution, ziatdinov18_build_explor_librar_atomic_defec_graph, rashidi20_deep_learn_guided_surfac_charac}; or classifying the state of particular atoms or molecules \citep{rashidi_2018_asm_in-situ_tip_coniditioning_through_ml}.
Yet others have begun reevaluating the overarching design of experiments around using statistics to drive the decision of what to do next during the experiment \citep{noack_2021_gaussian_processes,Kelley_2020_fast-spm-via-ml,thomas22_auton_scann_probe_micros_inves,liu_2022_hypothesis-driven_automation_spm}.
Active learning, where a machine learning algorithm's internal model is updated during the experiment, has been introduced outside of SPM to allow optimal efficiency in data acquisition \citep{noack_2021_gaussian_processes}, and been investigated within SPM both for 2D scan efficiency \citep{Kelley_2020_fast-spm-via-ml} as well as scanning tunneling spectroscopy \citep{thomas22_auton_scann_probe_micros_inves}.
Furthermore, hypothesis learning, where one uses machine learning to validate between a number of hypotheses by testing these during the experiment, has been investigated \citep{liu_2022_hypothesis-driven_automation_spm}.

A major limitation that has permeated released source code for many of these experiments is reusability.
In most cases, the source code is written for a specific SPM model, making it non-trivial to take the code and reuse it on SPMs from other manufacturers.
This is particularly endemic with automation, where the machine learning component is run within the experimental loop.
Even in cases where a researcher has designed their code to be as loosely coupled as possible, unnecessary additional dependencies and poor documentation often make reuse difficult.
We should highlight some work that has been done in this space: in Ref. \cite{vasudevan2023processing}, a software ecosystem was developed to simplify reading and analysing data from various scientific instruments (with one of the packages available at Ref. \cite{githubPycroscopySciFiReaders}).
While very useful, this does not solve the interfacing issues with reusing code for online experiments.
Another contribution was made in Ref. \cite{liu2024aecroscopy}, where a hardware-software framework was developed to allow automated experimentation.
This framework focuses on the design of low-level particulars, to give researchers finer control on aspects of an experiment (e.g. scanning trajectories and excitation waveforms).
This deeper granularity of control is quite useful, but requires usage of a separate hardware component.
We consider both of these developments complementary to our own, and in fact use Ref. \cite{githubPycroscopySciFiReaders} for some of the scan reading procedures in our framework.

With this in mind, we have developed an automation framework for SPM (that we have named \texttt{afspm}) aimed toward facilitating code sharing and reusability of developed components.
In this paper, we begin by overviewing key features and design characteristics of the framework.
Afterward, we present the materials and methods used to validate our approach and discuss the results obtained.
Finally, we conclude with potential further work to expand upon this tool.

\section{Framework Design Overview}

In what follows, we highlight key characteristics of our developed framework.
Note that more in-depth information is described in the supporting information.

\subsection{Main Design Characteristics}

Our framework concerns itself with automation of the high-level, low-frequency decisions a researcher would perform during the running of an experiment (e.g. where to scan, when to scan).
We seek to explicitly differentiate this from the low-level, high-frequency commands handled by an SPM controller (e.g., a feedback control loop); and the more involved steps performed during setup and teardown of an experiment, such as those needed to prepare a sample under ultra-high vacuum.

\begin{figure}
  \begin{subfigure}[b]{0.28\textwidth}
    \centering
    \resizebox{\textwidth}{!}{  
      \includegraphics[width=\textwidth]{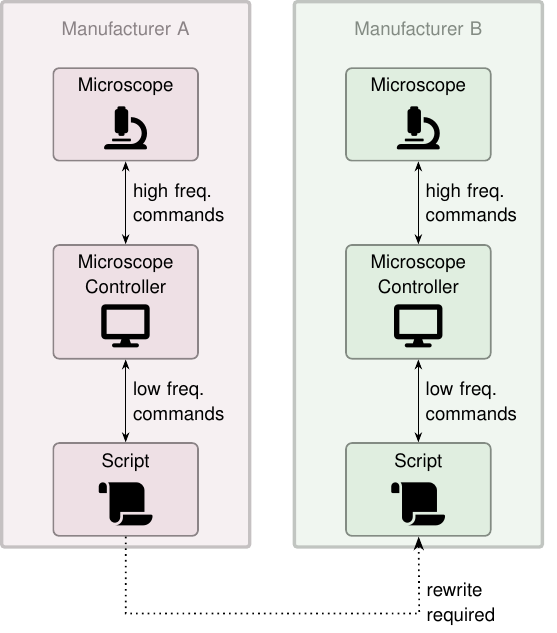}
    }
    \caption{}
    \label{fig:code_reuse_without_afspm}
  \end{subfigure}
  \begin{subfigure}[b]{0.32\textwidth}
    \centering
    \resizebox{\textwidth}{!}{  
      \includegraphics[width=\textwidth]{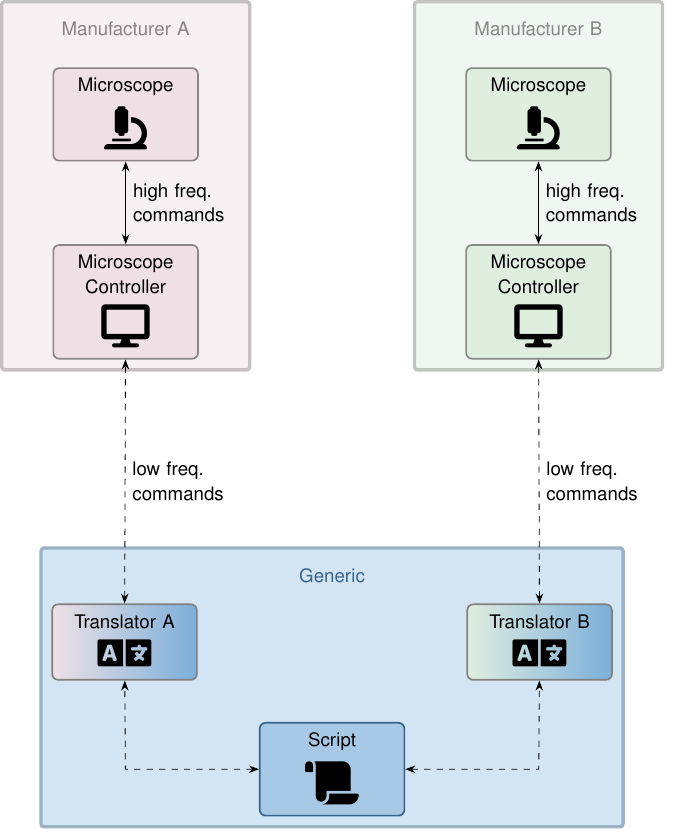}
    }
    \caption{}
    \label{fig:code_reuse_with_afspm}
  \end{subfigure}
  \begin{subfigure}[b]{0.32\textwidth}
    \centering
    \resizebox{\textwidth}{!}{  
      \includegraphics[width=\textwidth]{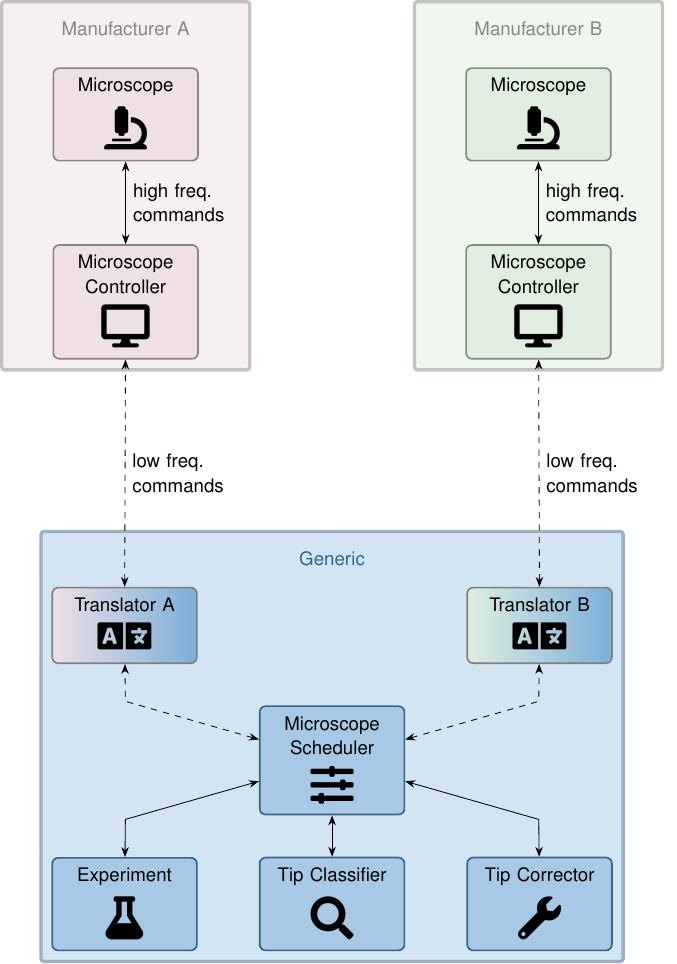}
    }
    \caption{}
    \label{fig:code_afspm_with_components}
  \end{subfigure}
  \caption{(a) Code reuse today, where experimental scripts are written for a particular microscope and reuse on a new microscope usually requires a full rewrite.
    (b) Usage of a similar script in \texttt{afspm} involves sending generic commands to a Microscope Translator, which translates these to manufacturer-specific commands.
    (c) Rather than a one-off experimental script, an experiment can be designed by hooking together various pre-developed automation components as in \texttt{afspm}, as with this exemplary set of components.
    The Microscope Scheduler ensures that only one of these components controls the microscope at any given point in time.}
  \label{fig:code_reuse}
\end{figure}

A general overview of our framework can be seen from Figure \ref{fig:code_reuse}.  
In Figure \ref{fig:code_reuse_without_afspm}, we see the difficulty with code reuse in SPM automation as it exists today.
When a piece of automation is written, it usually bundles both the automation logic and the commands sent to control the microscope into a single software file.
Unfortunately, there is no standard for SPM communication, meaning manufacturers use different command calls and in many cases involve different programming languages.
Because of this, reusing said automation on a microscope from a separate manufacturer usually requires a complete rewrite.
Going forward, we will refer to these one-off pieces of code as scripts, indicating their single-purpose nature.
In our framework, we aim to avoid these kinds of rewrites by decoupling the SPM-specific commands from the logic of the experiment we wish to run.
This is accomplished by defining commands and data structures in a generic format and allowing a manufacturer-specific translator to convert these into each manufacturer's specific commands and data structures when sending them out to a given controller (see Figure \ref{fig:code_reuse_with_afspm}).
A script written using these generic commands and schemas can then be run on any microscope, as long as a microscope translator for it exists.

In practice, much of the code that is written for an experiment is common.
This aligns with the reality that there are many common tasks which must be performed when running an SPM experiment (e.g. characterizing the tip, accounting for thermal drift).
The \texttt{afspm} package is designed around the concept of components, each responsible for an explicit purpose.
Here, a component refers to a standalone piece of code that has implicit required inputs to run and outputs to send out; it differs from a script in the assumption that it can be configured and run for various different experiments.
In Figure \ref{fig:code_afspm_with_components}, we see how the one-off script from Figure \ref{fig:code_reuse_with_afspm} was redefined around three exemplary components in \texttt{afspm}: an Experiment, which determines where to scan over the course of the experiment; a Tip Classifier, responsible for detecting if the tip morphology has changed appreciably; and a Tip Corrector, which will attempt to fix the tip if required.
We can imagine these components as the software equivalents to the hardware instruments associated with a given microscope: just as these hardware instruments are hooked together in a particular fashion to run the microscope, we connect various software components in order to run a desired experiment.
By properly subdividing our experiment into these standalone components and designing each with general usage in mind, we hope to encourage the development of reusable automation components that can be easily shared among the community.

\subsection{Experiment Configuration and Running Experiments}

\begin{figure}
  \begin{subfigure}[b]{0.25\textwidth}
    \centering
    \resizebox{\textwidth}{!}{  
      \includegraphics[width=\textwidth]{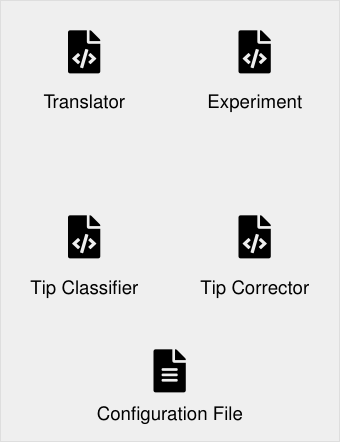}
    }
    \caption{}
    \label{fig:experiment_components}
  \end{subfigure}
  \begin{subfigure}[b]{0.32\textwidth}
    \centering
    \resizebox{\textwidth}{!}{  
      \includegraphics[width=\textwidth]{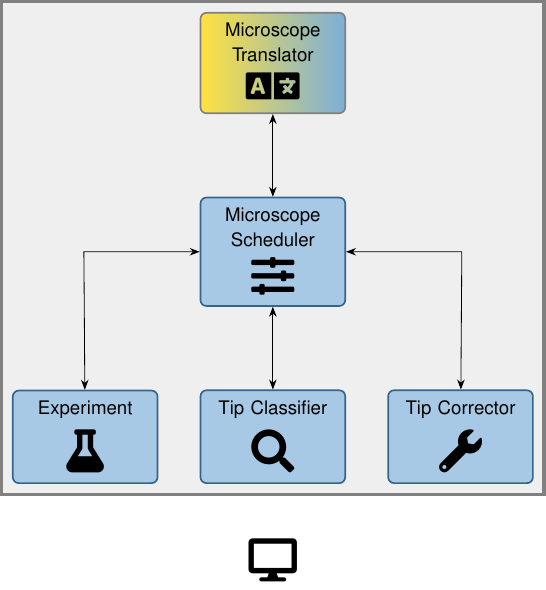}
    }
    \caption{}
    \label{fig:running_one_system}
  \end{subfigure}
  \begin{subfigure}[b]{0.32\textwidth}
    \centering
    \resizebox{\textwidth}{!}{  
      \includegraphics[width=\textwidth]{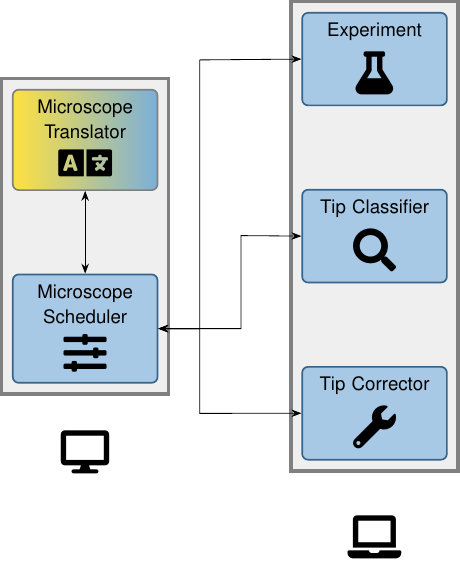}
    }
    \caption{}
    \label{fig:running_two_systems}
  \end{subfigure}
  \caption{(a) An automated experiment's configuration is defined via a configuration file, which defines the various software components to run, their parameters, and how they send data to each other. The various software components are pieces of code written in the \texttt{afspm} framework with the aim of being reusable. The software components here match the experiment configuration from Figure \ref{fig:code_reuse}. This experiment can be run all on one computer, as in (b), or split among multiple computers (e.g. to avoid computational bottlenecks) , as in (c).}
  \label{fig:config_and_running}
\end{figure}

Our framework uses a TOML-based configuration file \citep{tomlTOMLTomaposs} to define components of an experiment, their settings, and the manner by which they communicate (see Figure \ref{fig:config_and_running}).
When running a given experiment on a separate SPM, we expect and hope that only minor SPM-specific variables need to be changed (e.g. the scan region dimensions).
An automated experiment is begun via a command-line script, wherein the selected components are instantiated and monitored.
Components that crash or freeze are detected and restarted, allowing an experiment to continue under spurious crashes.
A user can instantiate individual components across multiple computers as desired.
In this way, a researcher may choose to split their experiment up across multiple computers in order to avoid memory or computational overhead issues or merely for personal convenience (see Figures \ref{fig:running_one_system} and \ref{fig:running_two_systems}).
While the existing components and tools are written in one programming language (Python), at its core the framework is defined solely by these generic commands and data structure schemas and the mechanism by which this data is sent to or from the different components.
In principle, a user can develop an automated component in the language of one's choosing, as long as it is able to access the expected input and output data over the network, read the input, and write the output.

\subsection{Common Components}

A given experiment may consist of any number of components, but every experiment will usually include two: a Microscope Translator and a Microscope Scheduler (see Figure \ref{fig:code_afspm_with_components}).
The translator is specific to a given SPM model or manufacturer (and thus requires an individual piece of code), while the scheduler is generic and is thus already written, only requiring that certain aspects be defined in an experiment's configuration.

The Microscope Translator serves as the interface between an SPM's specific programming interface and our generically defined \texttt{afspm} one.
It is responsible for responding to any requests sent to the SPM and publishing any changes that have occurred related to the SPM.
For example, a component in the experiment may request to change the SPM's physical scan region and digital scan characteristics; the SPM must attempt to perform the action and respond with either an indication of success or a response indicating why the action could not be performed.
If successful, the SPM would be expected to `publish' this change to anyone listening.
In attempting the action, the translator must convert the generic input command to the microscope specific command.

The Microscope Scheduler is responsible for mediating control to the SPM and caching data received so any recently connected component can function properly.
It thus sits between the Microscope Translator and all other components, scheduling control.
In order to avoid multiple components sending conflicting requests at once, \texttt{afspm} allows only a single component to be in control of the SPM at any point in time.
From a design standpoint, the mediation logic is quite simple: the first component to request control (with the appropriate parameters) is granted control, and no other component can obtain control until that component releases it.
This design juggles simplicity with both the cost of race conditions on this request, and the risk of a `greedy' component blocking the experiment from continuing.
Separately, a component not in control may request control and add/remove ‘flags’ indicating a detected problem with the experiment (for example, a tip
characterization issue).
By this mechanism, a component may pause regular running of the experiment: if one or more problems are flagged, the Scheduler removes the current component in control.
At this stage, only components indicating they resolve one of the currently logged problems may obtain control.

Regarding the caching of data: a given component will require certain expected inputs in order to compute and send an expected output; if a component is initiated at the wrong time, it may not receive all of these inputs.
To avoid this issue, we include logic for caching data as they are sent around the network, with the manner in which it is done determined by the experiment's configuration.

\subsection{Base SPM Actions and Parameters in \texttt{afspm}}

When defining this framework, we attempted to define some minimum set of commands and data structures that we believe form a ‘base’ set to run automation tasks.
Such a definition is arbitrary and debatable; however, the high volume of controls available in most SPM controllers make the task of creating a fully comprehensive framework impractical.
We chose to focus on the most common actions and parameters: configuring and running scans, configuring and running spectroscopies, and configuring the z-height feedback.
For scans, the configuration pertains to the physical region we are scanning and their data resolution.
For spectroscopies, the configuration allows defining the physical position where the spectroscopy is to be run.
For z-height feedback, one can control the setpoint value, the proportional and integral gain parameters, and whether it is on or off.
This level of control assumes the user has pre-configured their `operating mode' (e.g. amplitude-modulated AFM, frequency-modulated AFM), and spectroscopic mode (e.g. IV curves, force spectroscopy), which we believe is justifiable as part of the experimental setup.
We should note, however, that a user of \texttt{afspm} is not restricted to this subset of controls: generic actions and generic parameters can be added as needed.
One could define actions to swap between operating modes and spectroscopic modes, or add parameters for controlling other feedback loops of a given operating mode.


This base set of actions and parameters allows components to respond to full scans and spectroscopies once they have finished.
This has the advantage that a given Microscope Translator does not need to monitor data live as it is being recorded by the microscope: it only needs to read each saved scan or spectroscopy, convert it into the associated generic data structure, and send it out to components.
This simplifies the complexity of writing a given translator.
However, this decision is limiting for experiments where decisions may need to be performed in between scans.
We should note that this is not a design limitation: support for reading scan data as the scan is performed can be added if necessary.

\subsection{Control Transfer Example}

\begin{figure}
  \includegraphics[width=\textwidth]{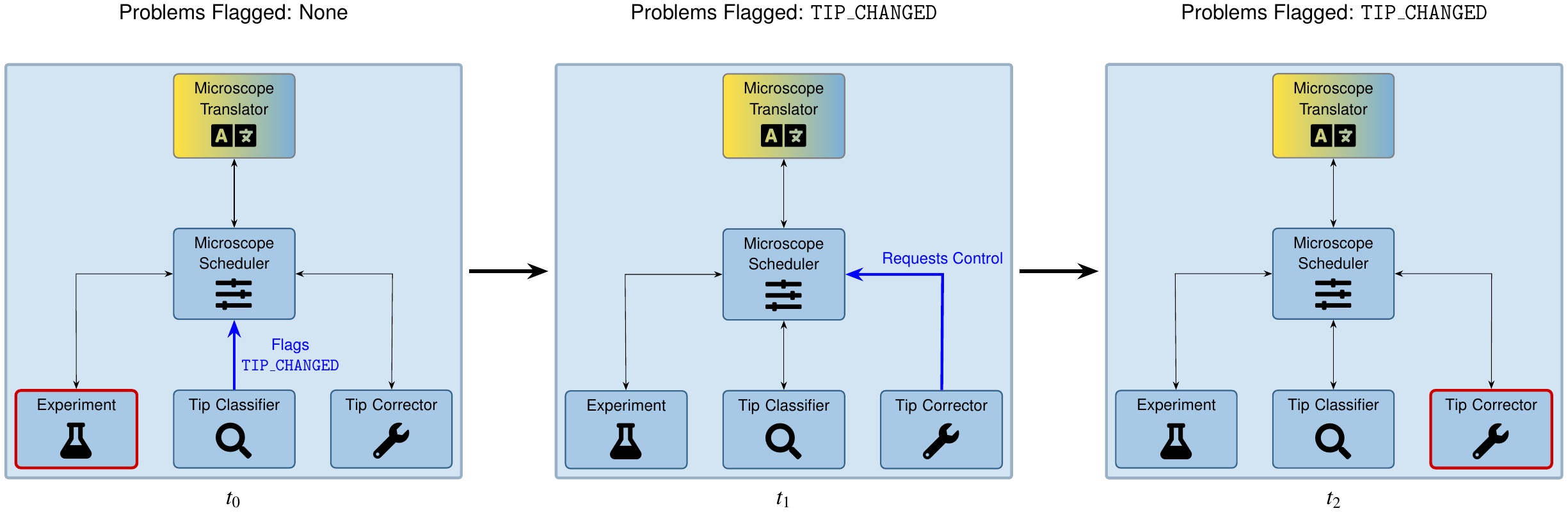}
  \caption{Example of control transition between two components in an experiment (with the component in control indicated by a red border and currently flagged problems on the top of each diagram).
    At time $t_{0}$, the Experiment component has been in control.
    The Tip Classifier, having detected that the tip has changed, flags an experiment problem.
    This flagging causes the Microscope Scheduler to take away control from the current component, waiting for either the problem to be removed, or a component to request control indicating it resolves said problem.
    At time $t_{1}$, the Tip Corrector component requests control, indicating it solves the tip problem.
    The Scheduler gives it control as seen at time $t_{2}$, and it can proceed to instructing the Microscope to perform actions in an attempt to fix the tip.}
  \label{fig:control_transition}
\end{figure}

We return to the experimental configuration of Figure \ref{fig:code_afspm_with_components} to illustrate how control is traditionally transferred between components.
This experimental design (consisting of an Experiment, a Tip Classifier, and a Tip Corrector) could apply, for example, to atomic resolution imaging with CO-functionalized tips.
In such an experiment, detecting that the tip morphology has changed and correcting for it is critical.
In Figure \ref{fig:control_transition}, we see how control is swapped between the three components of an example experiment over time (note that we indicate the component in control via a red border around it).
For the majority of the experiment, the Experiment component is in control.
This component performs the pre-defined experiment, such as scanning a series of images of the sample.
At $t_{0}$, the Tip Classifier component detects -- from analyzing the scans that have been made -- that the tip morphology has changed.
Accordingly, it flags an experiment problem indicating as much.
This causes the Microscope Scheduler to take away control from the Experiment component shortly before $t_{1}$.
At this point, no component is in control, and the Experiment component cannot regain control until the flagged problem is removed.
As the Tip Corrector is able to solve the tip morphology problem, it requests control at time $t_{1}$.
Since it resolves a currently flagged problem, the scheduler grants it control at $t_{2}$.
It will instruct the microscope to perform a series of actions in an attempt to fix the problem, such as pulsing the tip with a bias voltage or crashing into a different portion of the sample at a defined velocity.
Once the tip has been resolved, the Tip Classifier will detect this and tell the scheduler to remove the tip problem, causing the Microscope Scheduler to take away control from the Tip Corrector.
After this, the Experiment may request and regain control, continuing the experiment.

In this manner, many automation components can be running throughout an experiment, coordinating to resolve problems as they are detected while the experiment runs.
This decoupling of roles also simplifies writing, testing, and validating the code: the Experiment code only needs to go through the explicit scan regions it wishes to scan; the Tip Classifier only needs to analyze incoming scans and flag or unflag when it detects the tip has changed; and the Tip Corrector only concerns itself with attempting to correct the tip when the tip problem has been flagged.
We note that there is some ambiguity in terms of who should indicate the experiment problem has been resolved: depending on the approach taken, it may make more sense for the Tip Corrector to do so.
However, this ambiguity does not detract from the overall benefit of decoupling automation logic into function-specific components, as described here.

\section{Framework Validation}

We validated our framework by showing that our base set of SPM operations works as expected on two different SPMs and running an experiment using a drift correction component developed using \texttt{afspm}.

\subsection{Translator Test Suite}

In order to ensure a given Microscope Translator behaves as the framework expects, we developed an automated test suite that validates each of the base set of SPM operations.
Using this test suite validates that scripts that are written in a manufacturer-agnostic fashion can be run on a specific SPM.
The suite additionally lists all supported and unsupported generic actions and parameters, so one can ensure a given experiment is runnable by a translator.
This test suite was developed using \texttt{pytest} and can be easily invoked via command-line, allowing a user to run an automated sanity check before beginning an experiment.
Since these tests change parameters and perform various actions, we consider it a best practice to run them with the tip disengaged.
Although the created scans and spectroscopies will consist only of noise, this does not matter: the tests validate that the translator properly responds to requests, performs actions, and returns expected data types.
For example, we check that running a scan ultimately causes a new scan to be saved by the SPM controller, which our translator can then read, convert to our generic schema, and send out to be read by others.
The test suite exists alongside a comprehensive set of tests -- also developed using \texttt{pytest} -- that validate the framework for expected functionality.

\subsection{Generic Drift Correction}

Next, we aimed to show how reusable automation components could be developed by designing a generic drift correction component and running it during an experiment.
By drift, we refer to the thermal drift induced by the thermal expansion characteristics of the various components of the microscope.
These and their differences cause the relationship between the tip coordinate system (TCS) and the sample coordinate system (SCS) to vary over time, as temperature gradients expand or contract the materials associated to the sample and those of the probe containing the tip.
This variation makes it difficult to track the same regions of a sample continuously over time.
A drift correction component would be responsible for detecting this drift and correcting for it, such that other components can consider a single coordinate system and not concern themselves with this variation.
To explain our approach, we briefly highlght our algorithm for estimating drift and correcting for it, describe the overarching component workflow, and then discuss the division of this logic into components.


\subsubsection{Drift Correction Algorithm Overview}

Using an image registration technique for drift correction, we estimate linear transformations $M_{ij}$ between two scans $S_{i}$ and $S_{j}$, where $M_{ij}$ corresponds to a homogeneous linear transformation of the form $M = \left[ R_{\mathrm{3\times2}} \vert \bm{t}_{\mathrm{3\times1}} \right]$ containing both a $\mathrm{3\times2}$ rotation matrix $R$ and $\mathrm{3\times1}$ translation vector $\bm{t}$.
We track drift over time from the start of the experiment, considering a static SCS coordinate system and dynamic TCS one that is changing throughout the experiment.
We maintain a total mapping from TCS to SCS, denominated $M_{tot}$, which we initialize as $M_{tot} = \left[ I_{\mathrm{3\times2}} \vert \bm{0}_{\mathrm{3\times1}}\right]$ (i.e., no difference between SCS and TCS).
As the experiment runs, we grab time-subsequent scan pairs $\left(S_{i}, S_{j}\right)$ (chosen via logic explained below), and estimate the current drift $\widetilde{\Delta M}_{tot, k}$ between these, where $k$ is a scan step.
Simultaneously, we perform a first-order prediction $\widehat{\Delta M}_{tot, k}$ based on our prior drift estimates and use this to update our scan position when we begin a scan.
Our drift correction logic, then, involves the following steps:
\begin{equation}
  M_{tot,k}' = M_{tot,k-1} \cdot \widehat{\Delta M}_{tot, k}
  \label{eq:mapping_predictive_update_main}
\end{equation}
\begin{equation}
  M_{tot,k} = M_{tot,k}' \cdot \widetilde{\Delta M}_{tot, k}
  \label{eq:mapping_estimation_update_main}
\end{equation}
where the drift predictive step occurs in equation \eqref{eq:mapping_predictive_update_main}, used to set the scan region before performing a scan; and the estimation update occurs in equation \eqref{eq:mapping_estimation_update_main}, performed once the scan finishes.
By predicting drift at scan time and updating our predictions after the scan has finished, we attempt to reliably track drift throughout the experiment.
In doing so, we can encapsulate the drift estimation logic away from the other components: if the correction works sufficiently well, they only need to be aware of the SCS.

For estimating our drift mapping $M_{ij}$ between two scans $S_{i}$ and $S_{j}$, we use a feature detection approach similar to the approach taken in Ref. \cite{diao2023automatic} and that taken for aperiodic structures in Ref. \cite{dickbreder2023undrift}, where an algorithm detects common features in both scans and uses an outlier-robust mechanism to estimate the transformation necessary to align $S_{i}$ and $S_{j}$ spatially considering these feature pairs.
A more in-depth explanation of the drift estimation and correction algorithms used may be found in the supporting information.

\subsubsection{Overarching Workflow}

As our aim is to encapsulate drift correction logic, we have considered a passive design where we analyze scans as other components request them, looking for scan pairs with which we can estimate drift.
We only interfere with the existing experiment in extreme cases, when our estimated drift is too large (meaning that it has deviated too far from our linear prediction), and we believe a rescan of the last requested scan is necessary to avoid losing important portions of the scan region.
The definition of `too large' is experiment-dependent, defined by a scan deviation threshold ($f_{rescan}$, described below) that is set in the experiment configuration file.

To determine appropriate scan pairs, we search for prior scans that have a sufficient spatial intersection in the SCS and a sufficiently similar resolution.
We consider spatial intersection because our drift estimation algorithm requires common features in both scans and is not foolproof; in doing so, we limit unnecessary computation (for scan pairs where there is clearly no spatial intersection), and potential errors (if, for example, the drift estimation algorithm has incorrect feature matches between the two scans).
Spatial resolution is also considered to minimize feature matching errors, as the same feature in two scans may fail to be matched if the resolutions differ significantly.
We should note that many drift estimation algorithms aim to achieve scale invariance, potentially avoiding the need for this extra filter.

In an environment where the sample were unchanging and we are certain our measuring device does not change the sample, the best scan pairing approach would be to grab the oldest scan that meets our criteria throughout the lifetime of the experiment.
However, we did not want to impose this assumption on our workflow, and therefore consider a fixed-size running queue of prior scans, where the size of this queue is a configuration parameter.
One can make this queue very large for experiments where we have high confidence the sample is unchanging and smaller when we are less certain.
Note that the prior scans are added to the queue after having estimated drift at that time instance; as such, all scans in the queue should be relatively correctly positioned in the SCS.
However, we should note that this running queue risks accumulating errors, as our `ground truth' image is updated when the queue size limit is reached (causing us to remove the oldest scan, leaving the second oldest as our new ground truth).
Any error in our drift estimation between these two oldest scans is then added to our mapping, because our `ground truth' has changed.
We did not witness this causing problems in practice, but this limitation may need to be accounted for in the future.

\begin{figure}
  \centering
  \resizebox{0.75\textwidth}{!}{
    \includegraphics[width=\textwidth]{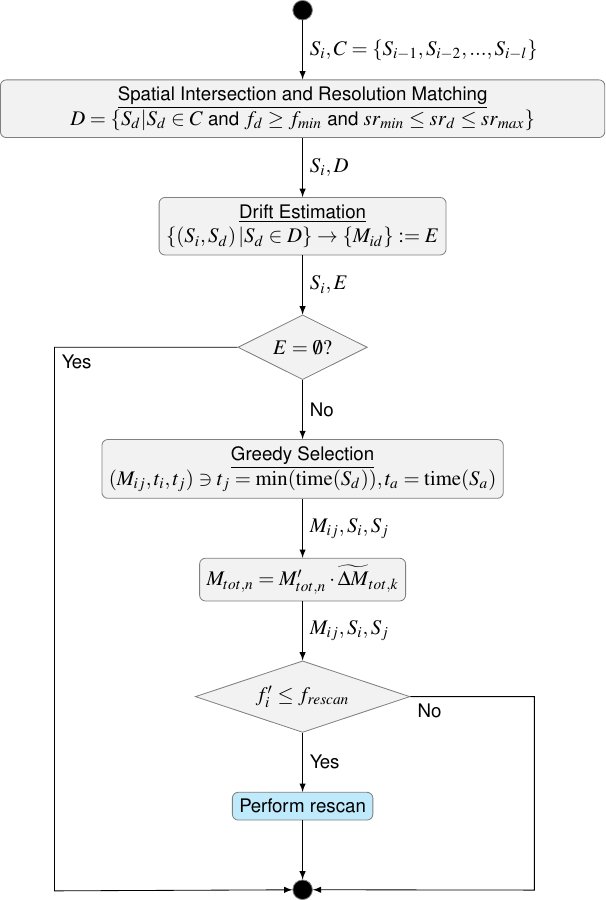}
  }
    \caption{The drift correction component updates its mapping $M_{tot}$ via the above workflow.
      Considering prior scans stored in a cache, we first filter out scans with insufficient spatial intersection or too dissimilar spatial resolution, resulting in a filtered set $D$.
      For each pair of $(S_{i}, S_{d})$, a mapping $M_{id}$ is estimated, and we select the scan pair with the oldest prior scan.
      If a mapping exists from this, we can update the mapping using equation \eqref{eq:mapping_estimation_update_main}.
      We also determine whether a rescan is warranted, for cases where the difference between our predicted drift $\widehat{\Delta M}_{tot, k}$ and estimated drift $\widetilde{\Delta M}_{tot,k}$ are too dissimilar.}

  \label{fig:overarching_correction_workflow}
\end{figure}

The overarching correction workflow is illustrated in Figure \ref{fig:overarching_correction_workflow}.
Throughout the experiment, we maintain a first-in first-out (FIFO) queue of the prior $l$ unfiltered scans, which we denote as the set $C$.
Assuming a function to extract the bounding rectangle of a scan:
$\mathrm{rect}\left(S_{i}\right) = \left(\bm{x}_{0}, \bm{x}_{1}\right) = \bm{r}_{i}$ -- with $\bm{x}_{0}$ the top-left position of the rectangle and $\bm{x}_{1}$ its bottom-right position -- one can estimate the intersection ratio $f_{a}$ of two such rectangles
$f_{a} = \mathrm{intersection\_ratio}\left(\bm{r}_{a}, \bm{r}_{i}\right) = \left. \bm{r}_{a} \cap \bm{r}_{i} \right/ \mathrm{min}\left(\bm{r}_{a}, \bm{r}_{i}\right)$.
Note that here and going forward we will use the subscript $a$ as a placeholder when defining functions; for example, $f_{j}$ would represent the result of $\mathrm{intersection\_ratio}\left(\bm{r}_{j}, \bm{r}_{i}\right)$.
In this case, the ratio is computed relative to the current scan $S_{i}$ and we can define a threshold $f_{min}$ for the lowest acceptable intersection ratio to consider a pair of scans for drift estimation.
For spatial resolution analysis, we refer to the ratio of data resolution of the scan over its physical size: ${sr}_{i} = \mathrm{spatial\_resolution}\left(S_{i}\right) = \left. \mathrm{resolution}\left(S_{i}\right) \right/ \vert \bm{x}_{1,i} - \bm{x}_{0,i}\vert$, where we can define acceptable bounds $sr_{min}$ and $sr_{max}$ in spatial resolution for prior scans $S_{c} \in C$ based on the spatial resolution of the current scan $S_{i}$.
Given the above, we can define our set of filtered scans $D$:
\begin{equation}
  D = \{ S_{d} \vert S_{d} \in C \hquad \mathrm{ and } \hquad f_{d} \geq f_{min} \hquad \mathrm{ and } \hquad {sr}_{min} \leq {sr}_{d} \leq {sr}_{max} \}
\end{equation}

If we have any prior scans in our filtered set $D$, we proceed to create a set of scan pairs $ \left( S_{i}, S_{d}\right) $, for which drift estimations are performed, leading to a set $E$ of estimates $M_{id}$.
In our current implementation we use a greedy algorithm: specifically, we only consider the drift estimate of the oldest prior scan for which the estimation succeeds:
\begin{equation}
  \left( M_{ij}, t_i, t_j \right) \ni t_j = \mathrm{min}\left(\mathrm{time}\left(S_d\right)\right)
\end{equation}
where $t_a = \mathrm{time}\left(S_a\right)$, and $\mathrm{time}\left(\cdot\right)$ returns the timestamp of a scan, such that $t_{a}$ is the timestamp of scan $S_{a}$.

If we have a successful drift estimation $M_{ij}$, we proceed to update our total mapping $M_{tot}$ using equation \eqref{eq:mapping_estimation_update_main} (where $\widetilde{\Delta M}_{tot, k}$ acts as a proxy for $M_{ij}$ in this scheme).
If we have not managed to estimate drift relative to any of the prior scans in our queue (i.e., $E = \emptyset$, our set of drift estimates $E$ is empty), we do not update our estimation for this step (no translation or rotation update beyond our predictive step).
Finally, we perform a comparison of the drift by considering the intersection ratio of $\bm{r_{i}}$ with $\bm{r}_{i}'$, its updated bounding rectangle: $f_{i}' = \mathrm{intersection\_ratio}\left(\bm{r}_{i}', \bm{r}_{i}\right)$.
If it is below a predefined threshold $f_{rescan}$, we deem that our scan has drifted too much and perform a rescan with $\bm{r_{i}}'$.
Rescanning therefore involves requesting a scan of the same region in the SCS, for situations where the drift estimate was sufficiently off from the detected drift, causing the scan to contain a proportion of the desired region that is too low.

\subsubsection{Component Design}

\begin{figure}
  \begin{subfigure}[b]{0.45\textwidth}
    \centering
    \resizebox{\textwidth}{!}{
      \includegraphics[width=\textwidth]{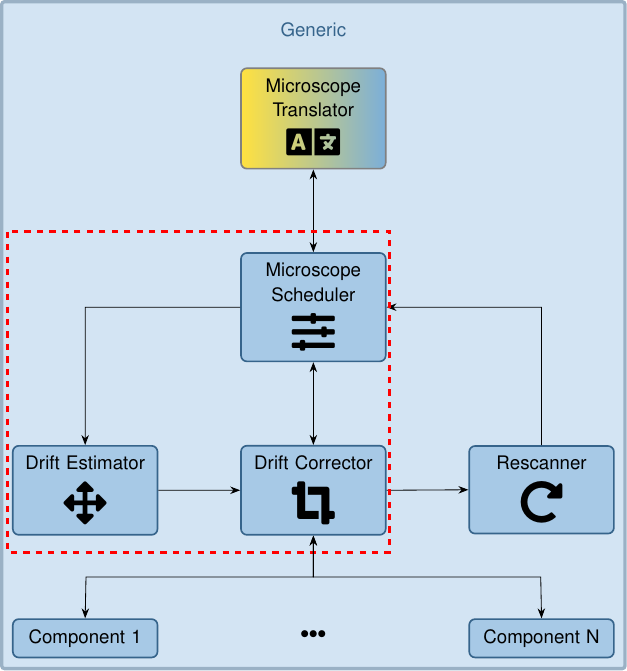}
    }
    \caption{}
    \label{fig:component_design_total}
  \end{subfigure}
  \hfill
  \begin{subfigure}[b]{0.45\textwidth}
    \centering
    \resizebox{\textwidth}{!}{
      \includegraphics[width=\textwidth]{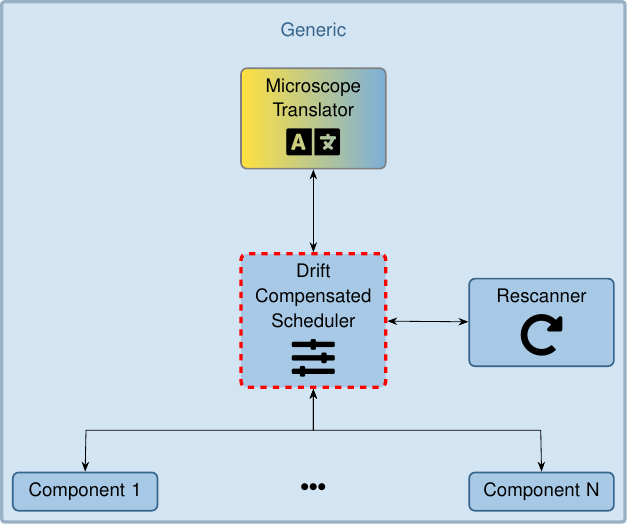}
    }
    \caption{}
    \label{fig:component_design_chosen}
  \end{subfigure}

  \caption{Illustration of the component design chosen for drift correction.
    In (a), we see the three drift correction functions subdivided into individual components: a Drift Estimator, to estimate drift between scans; a Drift Corrector, which receives these estimates and functions as a `pass-through' between components and scheduler (converting data structures sent out by the microscope from TCS to SCS, and requests to the microscope from SCS to TCS); and a Rescanner, which requests a rescan whenever the Drift Corrector has determined it necessary.
    In (b), we see our chosen design, where the logic of the Drift Estimator, Drift Corrector, and Microscope Scheduler are combined into a single component, denominated a Drift Compensated Scheduler (with its border dashed and red and the components it replaces bounded by a dashed red rectangle in (a)).
  }
  \label{fig:correction_component_design}
\end{figure}

Now, we must decide how to divide our drift compensation logic into one or more components.
In most cases, the division of automation logic into components is straightforward, as each discrete function we desire can be accomplished by an individual component.
Usually, all components will reside below the Microscope Scheduler, performing their individual functions while using the Scheduler to communicate with the microscope.
As we will see (and illustrated in Figure \ref{fig:correction_component_design}), drift correction is somewhat more complicated, introducing some degree of choice in how we subdivide the algorithm.

In the case of drift correction, we wish to: (i) track drift over time; (ii) output `drift corrected' data structures (e.g., scans and spectroscopies) to other components; and (iii) perform rescans of the latest requested scan if deemed necessary.
The second criteria is key here: if we desire to allow other components to think solely in the SCS, whatever component serves this function would become the main `hub' for microscope data.
All other components would connect to this component to receive data structures from the microscope in the SCS (converted from TCS to SCS using the mapping $M_{tot}$).
Similarly, any requests sent to the microscope would need to be converted from SCS to TCS (using the inverse mapping, $M_{tot}^{-1}$).
We can see an example view of independent components designed for these three purposes in Figure \ref{fig:component_design_total}.
Here, a Drift Estimator receives scans from the Microscope Scheduler, estimating drift and sending these drift estimates to the Drift Corrector.
The Drift Corrector component functions as this `main' hub, converting data between TCS and SCS as needed.
The Drift Corrector also compares the drift before and after a scan, determining if there was too much error in the drift prediction.
In this case, it sends a request to the Rescanner, indicating a rescan is required.
The Rescanner component is in charge of forcing these rescans when necessary.

The above described division of components, while reasonably well encapsulated, introduces complexity and confusion on the experiment design.
The Drift Corrector, in its role as main hub, is replacing the traditional role of the Microscope Scheduler.
Now, we have two stages of intermediation: one for scheduling control (performed by the Microscope Scheduler), and one for transforming data between SCS and TCS (performed by the Drift Corrector).
This extra stage creates extra complexity: we now have a stage of components that function purely in the TCS (the Microscope Translator and Microscope Scheduler);
a stage of components which are aware of both the TCS and SCS coordinate systems (the Drift Estimator, Drift Corrector, and Rescanner); and a stage of components that function purely in the SCS (all other components for an experiment).

Because of this extra complexity and confusion in roles, we decided to encapsulate the majority of the drift correction logic into an alternative Microscope Scheduler, which we name the Drift Compensated Scheduler (Figure \ref{fig:component_design_chosen}).
This scheduler performs the same functions as the Microscope Scheduler, the Drift Estimator, and the Drift Corrector.
We presume most users who wish to have drift compensation running would find it easier to simply replace the existing Scheduler with a drift-compensated one.
Note, however, that in our design the Rescanner remains a separate component.
We chose this because we found it confusing that the Scheduler could itself be in control (necessary to send rescan requests).


We should highlight that our division of components is a trade-off and -- like most decisions on encapsulation -- there is no single optimal design.
We acknowledge the community may decide it prefers a different encapsulation, which we may accordingly switch to.
We expect that over time these choices will iterate to more optimal divisions of components.
We also note that this encapsulation choice does not significantly limit user configuration: the chosen drift estimation algorithm is currently a configuration option in the Drift Compensated Scheduler.
If a researcher desired to use a different algorithm (such as measuring drift via the periodicity of the surface, assuming atomic resolution), they could add this as an option to the existing component.

\section{Results and Discussion}

\subsection{Validating Microscope Translation}

We developed translators for the Gnome X Scanning Microscopy project (GXSM) \citep{zahl2015gxsm}, an open-source software and hardware controller system, as well as an Asylum Research system.
In the GXSM case, our hardware controller was the softdB Signal Ranger MK2-810, and we were able to interface to the GXSM software using the pre-existing Python-based programming interface.
Minor changes were made to GXSM's Python programming interface to simplify canceling and restarting scripts, pushed to GXSM's main development branch.
Since a pre-existing Python package for reading GXSM scans did not exist, one was written called \texttt{gxsmread} \citep{githubGrutterSpmgGoupGxsmread}.
In the Asylum Research case, the specific controller was the MFP-3D-BIO, which uses an IGOR-based programming interface.
To deal with some interfacing particularities, our AsylumTranslator converts generic commands to SPM-specific commands serialized via JSON and transferred to the controller software via \texttt{ZeroMQ} \citep{zeromqZeroMQ}.
This extra interface is provided by a Python package called \texttt{ZeroMQ-XOP} \citep{zeromqXOPZeroMQXOP}.
We made minor modifications to this tool in order to support the underlying data analysis software used by Asylum Research --- an earlier version of Igor PRO.
These changes were pushed to \texttt{ZeroMQ-XOP}'s main development branch.
For reading Asylum Research scans, the Python package \texttt{SciFiReaders} was used \citep{githubPycroscopySciFiReaders}.
In both cases, the programming interface is common to all the manufacturer's microscopes, so we expect each translator to be reusable for other hardware from the same manufacturer.

Using the translator test suite, we were able to validate that both the GXSM translator and the Asylum translator supported all of our base SPM operations.
To run the translator test suite on the GXSM system, we simply disconnected the outputs of the SPM controller (softdB Signal Ranger MK2-810).
To run on the Asylum system, we flipped the system scan head.
This enabled us to test the system flow for both without risking affecting the rest of the system.
Videos showing both of these in action can be found in the supporting information.
This validates that experiments can be written in our generic framework and run on different SPMs.

\subsection{Validating Drift Correction}

In order to validate the functioning of our drift correction component, we ran experiments with it on the Asylum Research MFP-3D-BIO atomic force microscopy (AFM) system, scanning a CD stamper as a test sample using a MikroMasch HQ:NSC15/Al AFM probe (40 N/m force constant, 325 kHz resonant frequency).
These experiments were run in ambient conditions in an amplitude modulating (AM-AFM) mode.
Prior to engaging with the sample, we performed an auto-tuning operation to optimize the cantilever frequency, and manually optimized the z-height feedback parameters for reasonable imaging quality.
In the experiment, we attempted to scan the same of our sample over a long period of time ($\sim$ 16 hours).
We modeled the drift as 2D translations and thus dealt with a mapping $M_{tot} = \left[I_{\mathrm{3\times2}} \vert \bm{x}_{tot} \right]$, and thus focus only on the translation $\bm{x}_{tot}$ (where here we use $\bm{x}$ instead of $\bm{t}$ to avoid confusion with $t$ which we have used to designate time).
For our drift estimation, we chose to track a channel representative of the topographic variation on the sample.
For this operating mode and microscope setup, this corresponded to the `Height Retrace' channel.
Note also that we used the instrument's internal `Flatten1' scan pre-processing step, which fits an average line to each scanline and removes it.

\begin{figure}
  \centering
  \resizebox{0.5\textwidth}{!}{
    \includegraphics[width=\textwidth]{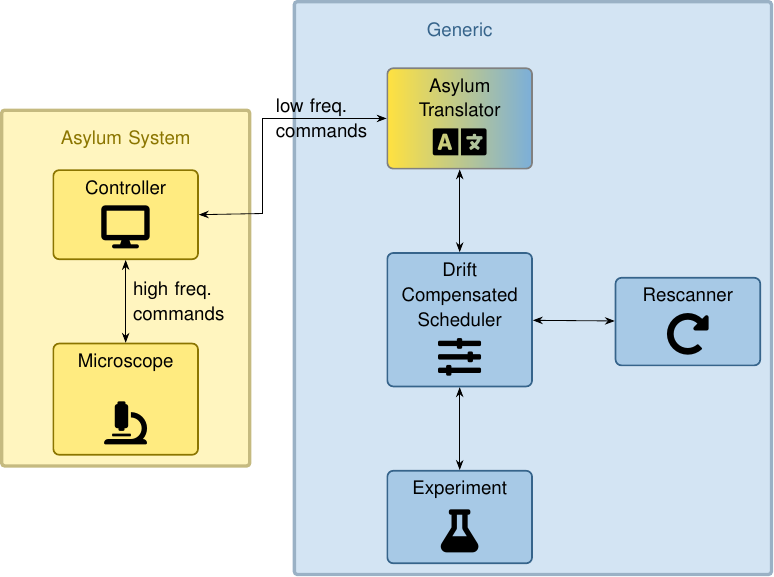}
  }
  \caption{Configuration of our automation experiment connecting to an Asylum system via our Asylum translator.
    In it, the Drift Compensated Scheduler handles both microscope scheduling and drift correction, sending drift compensated data structures out to the connected components.
    Linked to it is the Rescanner, which requests rescans if the system has drifted too far from the desired region.
    The Experiment component defines where to scan and at what frequency, in this case scanning a constant region every 15 minutes.
    }
  \label{fig:experiment_configuration}
\end{figure}

The configuration of our experiment is visualized in Figure \ref{fig:experiment_configuration}.
In it, we see that our configuration file specifies an Asylum translator to communicate with the Asylum controller.
Our drift correction logic is embedded within the special Drift Compensated Scheduler component, and performs the logic described above, converting all published data structures into ones in the Sample Coordinate System (SCS), and converting all data structures involved in requests to the microscope into the Tip Coordinate System (TCS).
In doing so, it allows the experiment designers to think in a single coordinate system.
Linked to it is a separate Rescanner component, that receives rescan regions from the Drift Compensated Scheduler when the drift between scans exceeds the configurable threshold $f_{rescan}$.
The Rescanner will then send those requests to the microscope (via the scheduler).
The Experiment component is an experiment-specific component written to scan a constant region at a given scan frequency.
We should note some additional components not visualized, that are useful for most experiments: a Control UI component provides a user interface such that the user can pause the automated experiment (to take control) or end the experiment.
A Metadata component saves useful information related to the various scans and spectroscopies captured that is not found in the saved files themselves (such as who was in control of the scan when it was performed).
This permits easier filtering of scan and spectroscopy data after collection.

We decoupled our drift correction experiments into two phases: scanning with our correction disabled and with it enabled.
By scanning with our correction disabled, we were able to roughly characterize the thermal drift.
This is useful in order to ensure our experiment configuration and drift correction parameters are reasonable.
Since our drift estimation algorithm relies on keypoints being matched between two scans, we desire that (a) the region we are scanning has sufficient surface variation to detect keypoints; and (b) the drift is detectable by our scans, both in terms of spatial resolution and scanning frequency.
The experimental parameters that may affect this are: the scanning resolution, the physical scan size, and the scanning frequency.
There are additionally somewhat indirect drift correction parameters to be considered: the acceptable spatial region intersection amount defining $f_{min}$; the acceptable spatial resolution deviations that define ${sr}_{min}$ and ${sr}_{max}$; the acceptable intersection amount before which we force a rescan, $f_{rescan}$; and any averaging added to our drift estimates between scans.
The chosen experimental and drift correction parameters for our experiment can be found in Table \ref{table:chosen_parameters}.
For the experimental parameters, we chose values that seemed reasonable following our experimental run with no correction.
For the drift correction parameters, we kept system defaults which were determined from minimal testing during the development of this component.
We should note that further evaluation to determine optimal defaults should be performed.
Note that in our code, rather than defining ${sr}_{min}$ and ${sr}_{max}$, we use ${sr}_{norm} = \left. \mathrm{min}\left({sr}_{i}, {sr}_{j}\right) \right/ \mathrm{max}\left({sr}_{i}, {sr}_{j}\right)$, and check if ${sr}_{norm} \leq {sr}_{norm, min}$.
This serves the same function as our original explanation.
Also, $w_{update}$ was not described prior: it is a weight factor applied to the currently estimated $\Delta M_{tot, n}$, to be averaged with $\widehat{\Delta M}_{tot, n}$.
By default, we do no averaging and use the latest estimate every scan.
Regarding the correction parameters, we expect $sr_{norm, min}$ and $f_{min}$ to be relatively independent of experiment run and keypoints estimation method used.
However, $f_{rescan}$ and $w_{update}$ may benefit from adjustment dependent on the desires and expectations of the experiment.
For example, $f_{rescan}$ may be set higher if there are stricter requirements on how much of the scan region must be within every scan over the duration of the experiment.

\begin{table}
  \centering
  \begin{tabular}{llr}
    \toprule
    \textbf{Category} & \textbf{Parameter} & \textbf{Value} \\
    \midrule
    Experimental & Scan Resolution [pix] & (128, 128) \\
    & Scan Region [$\mu$m] & (2.0, 2.0) \\
    & Scan Frequency [min] & 15 \\
    Drift Correction & ${sr}_{norm, min}$ &  0.25 \\
    & $f_{min}$ & 0.5 \\
    & $f_{rescan}$ & 0.75 \\
    & $w_{update}$ & 1.0 \\
    \bottomrule
  \end{tabular}
  \caption{Experimental and drift correction parameters used during the experiment.
    The experimental parameters were determined heuristically during the experimental run with correction off.
    The correction parameters are defaults determined from minimal experimentation during component development.}
  \label{table:chosen_parameters}
\end{table}

\begin{figure}
  \centering
  \includegraphics[width=\textwidth]{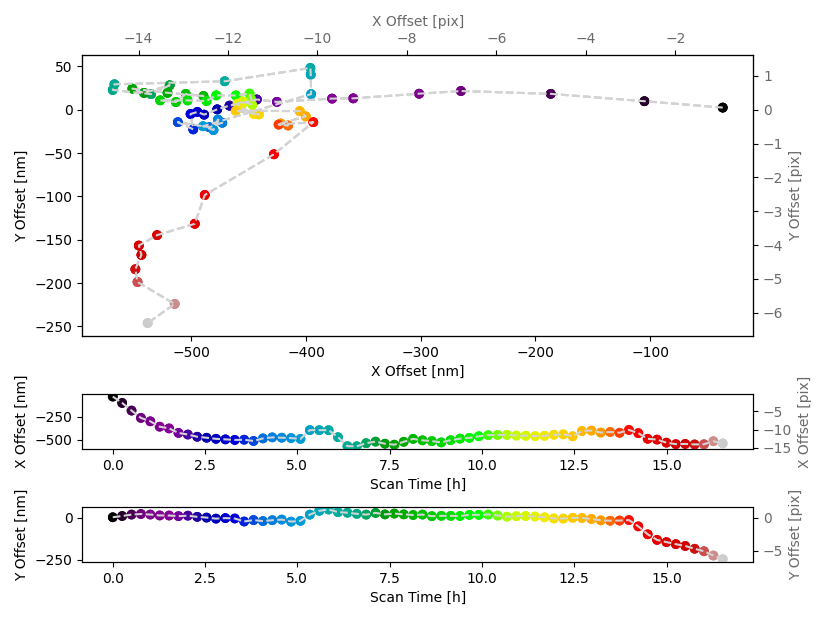}
  \caption{Drift offset $\bm{x}_{tot}$ tracked over time during the experiment (i.e. online), with drift correction enabled.
    The top graph shows the drift offset in x and y dimensions, with the bottom two graphs showing the individual dimensions on one axis and time on the other.
    The colour coding of the data points in all three graphs indicates time.}
  \label{fig:drift_correction_results}
\end{figure}

For our test, we requested to scan a specific region over $\sim$2 hours with the correction logic off.
For a second experimental run, we again scanned a specific region over $\sim$16 hours, this time with the correction logic on.
To analyze the data, we visualized the measured drift offset $\bm{x}_{tot}$ for the case where correction was on in Figure \ref{fig:drift_correction_results}.
Here, we see that the tracked drift follows a non-linear trajectory, similar to the results in Ref. \citep{diao2023automatic}.
In contrast with that paper, we did not evaluate the ultimate precision and stability of our corrector; rather, we focused on validating that it tracked correctly and minimized drift between scans.
Further analysis validating the correction can be found in the supporting information.
In Figure \ref{fig:experiment_scan_pairs}, we show some sample scan pairs for both experimental runs, roughly 2 hours apart.
In (a), we see scans with the drift correction off, while (b) show scans with the correction on.
Here, we can confirm visually that the drift was largely corrected for.

\newcolumntype{Y}{>{\centering\arraybackslash}X}
\begin{figure}
  \begin{tabularx}{\linewidth}{YY}
    \textit{\textbf{No Drift Correction}} & \textit{\textbf{With Drift Correction}} \\
    \begin{subfigure}[b]{0.4\textwidth}
      {
        \begin{tabularx}{\linewidth}{Y}
          \includegraphics[width=\textwidth]{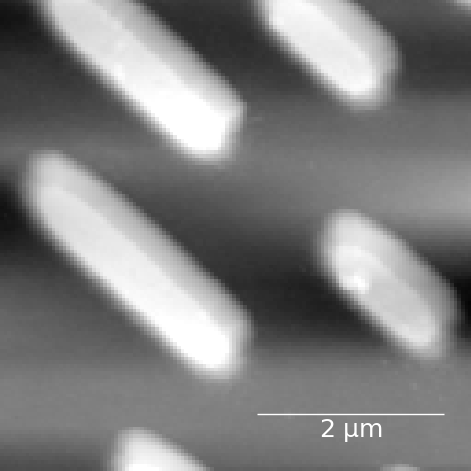} \\
          \centering{\textit{t}} \\
          \vspace{1em}
          \includegraphics[width=\textwidth]{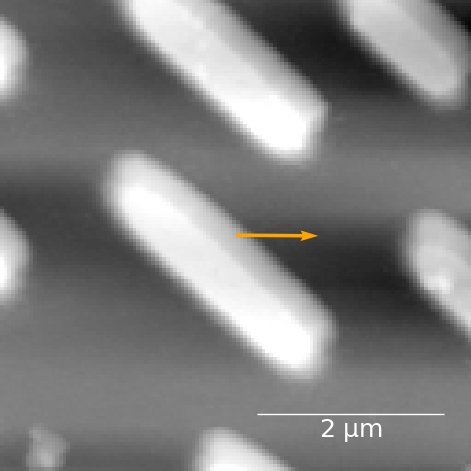} \\
          \centering{\textit{t + 2h}} \\
          \caption{}
        \end{tabularx}
      }
    \end{subfigure}
    &
      \begin{subfigure}[b]{0.4\textwidth}
        {
          \begin{tabularx}{\linewidth}{Y}
            \includegraphics[width=\textwidth]{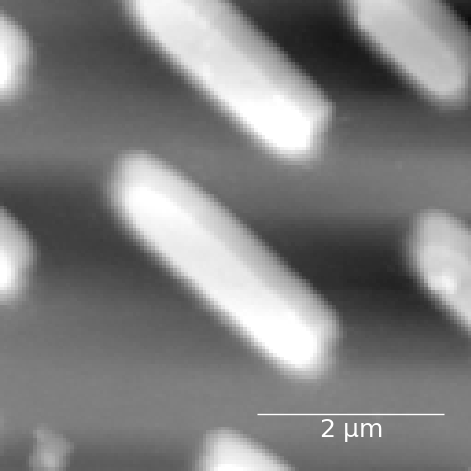} \\
            \centering{\textit{t}} \\
            \vspace{1em}
            \includegraphics[width=\textwidth]{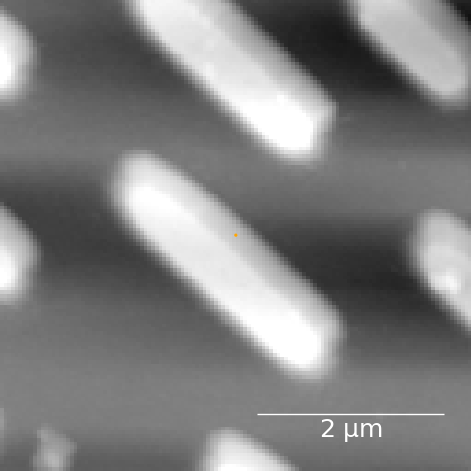} \\
            \centering{\textit{t + 2h}} \\
            \caption{}
          \end{tabularx}
        }
      \end{subfigure}
  \end{tabularx}
  \caption{(a) Two sample scans from the experimental run with correction off.
    The orange vector in the second scan (bottom-left) indicates the detected translation vector drift between scans.
    (b) Two sample scans from the run with correction on.
    In this case, the detector translation vector is so low as to be only a pixel in size.
    In both cases, the scan pairs are roughly 2 hours apart.}
  \label{fig:experiment_scan_pairs}
\end{figure}

\section{Conclusions and Outlook}

We present \texttt{afspm}, a software framework for writing automated experiments in scanning probe microscopy.
The framework defines generic commands and data structures for running experiments, translating to a microscope's specific commands and structure only as it is sent to the microscope controller.
By encapsulating function-specific code into components that communicate with each other, \texttt{afspm} encourages the development of automation code that is reusable.
The framework additionally benefits from allowing experiments to be split up among multiple computers, allowing computationally expensive logic (e.g., some machine learning computation) to be run on dedicated computers.
Its design allows new components to be written in programming languages other than Python, providing extra flexibility to researchers and sidestepping a potential roadblock for running on some SPMs.
We also highlight that the design considerations discussed here (and elaborated in the supplementary info) are not limited to SPM; such a design could be repurposed for optical or electron microscopy with similar automation benefits.

We validated the ability to write and run generic code on multiple microscopes by testing the functionality of translators written for two different microscopes.
We demonstrated writing a reusable component by developing a component for drift correction and confirmed its functionality by running a long-duration experiment tracking a constant region on a sample.

The development of automated components within a manufacturer-agnostic framework allows for sharing of code among the scanning probe microscopy community.
We consider this framework a reasonable first step toward developing generic automation in SPM.
As more automation components are developed and integrated, more useful and compelling experiments will become accessible to the community at large.
We hope \texttt{afspm} encourages this collaboration and would be delighted to assist others in writing reusable automated components.

Our developed framework is available from the Grutter Group Github and can be accessed at Ref. \cite{githubGrutterSpmgGoupAfspm}.



\section{Funding}

The authors acknowledge NSERC, INTRIQ, RQMP and the NRC for funding.


\nocite{protobufProtocolBuffers}
\nocite{zeromqZeroMQ}
\nocite{tomlTOMLTomaposs}

\nocite{diao2023automatic}
\nocite{dickbreder2023undrift}
\nocite{lowe1999object}
\nocite{bay2006surf}
\nocite{calonder2010brief}
\nocite{rublee2011orb}
\nocite{alcantarilla2012kaze}
\nocite{alcantarilla2011fast}
\nocite{derpanis2010overview}

\nocite{freedman1981histogram}

\newpage
\bibliography{nsullivan_afspm}

\pagebreak
\begin{center}
  \textbf{\Large Supporting Information \\ for \\ \texttt{afspm}: A Framework for Manufacturer-Agnostic Automation in Scanning Probe Microscopy}
\end{center}

\begin{appendix}
 \section{Design Particulars}
\subsection{Communication Protocol}
\subsubsection{Overview}

\begin{figure}
  \centering
  \resizebox{0.75\textwidth}{!}{
  \includegraphics[width=\textwidth]{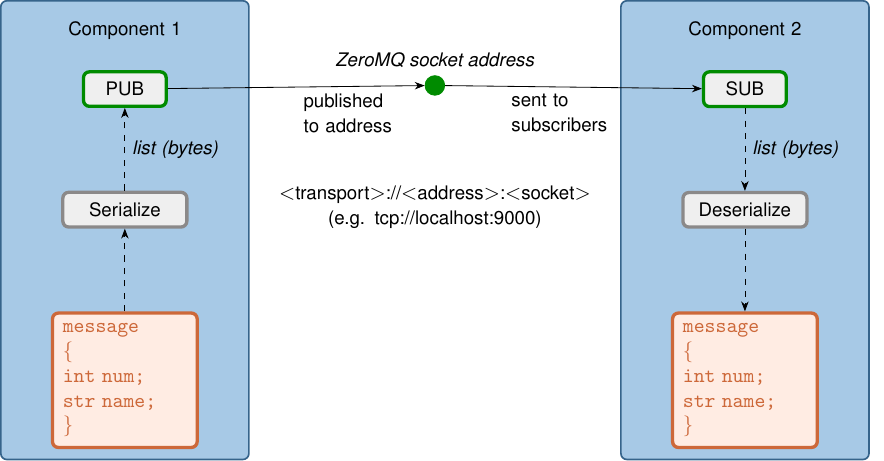}
  }
  \caption{The communication protocol in afspm involves using \texttt{Protocol Buffers} to serialize data structures into a sequence of bytes (and the reverse deserialization process); and using \texttt{ZeroMQ} to send data between defined software components.}
  \label{fig:communication_protocol}
\end{figure}

Our framework is defined by predefined commands and data structure schemas which are sent to and from network sockets in order to communicate between components (see Figure \ref{fig:communication_protocol}).
In order to be transmitted across a medium, it is necessary to have a serialization method: a process that translates the schema into a sequence of bytes that may then be reconstructed on the other end.
We have chosen Google \texttt{Protocol Buffers} \citep{protobufProtocolBuffers} as our serialization library, since it has multi-language/multi-platform support, guarantees type safety, and avoids schema violations.
The wide number of languages supported and ability to run on all major platforms ensures that a component may be written in almost any programming language.
By explicitly requiring type definitions when defining a schema, we avoid errors tied to incorrectly typed data being sent across the medium.
Similarly, the process of creating a data structure using \texttt{Protocol Buffers} ensures that the data sent adheres to the schema, avoiding extra data validation when reading a received schema.

To transmit our data between components we use \texttt{ZeroMQ} \citep{zeromqZeroMQ}, a cross-language messaging library that simplifies the process of sending data between applications.
\texttt{ZeroMQ} allows easy switching between in-process, inter-process, or network transports by changing the prefix of each socket address when created.
This makes it easy to switch between all components communicating on a single computer, each component on their own computer, or any combination in between --- it only requires changing the socket addresses and expliciting where components are created.
Additionally, a number of network communication roadblocks are handled automatically, such as reconnection, network synchronization, and message queueing issues.
It is a well-understood but perhaps rarely acknowledged fact that the development of a communication protocol for a given application imposes a significant workload on most developers' time; we hope our usage of \texttt{ZeroMQ} will sidestep such issues, allowing the researcher to focus on the experiment they wish to run over the intricacies of transferring data between components.

\subsubsection{Our publisher-subscriber, control-request world}

\begin{figure}
  \centering
  \resizebox{\textwidth}{!}{
    \includegraphics[width=\textwidth]{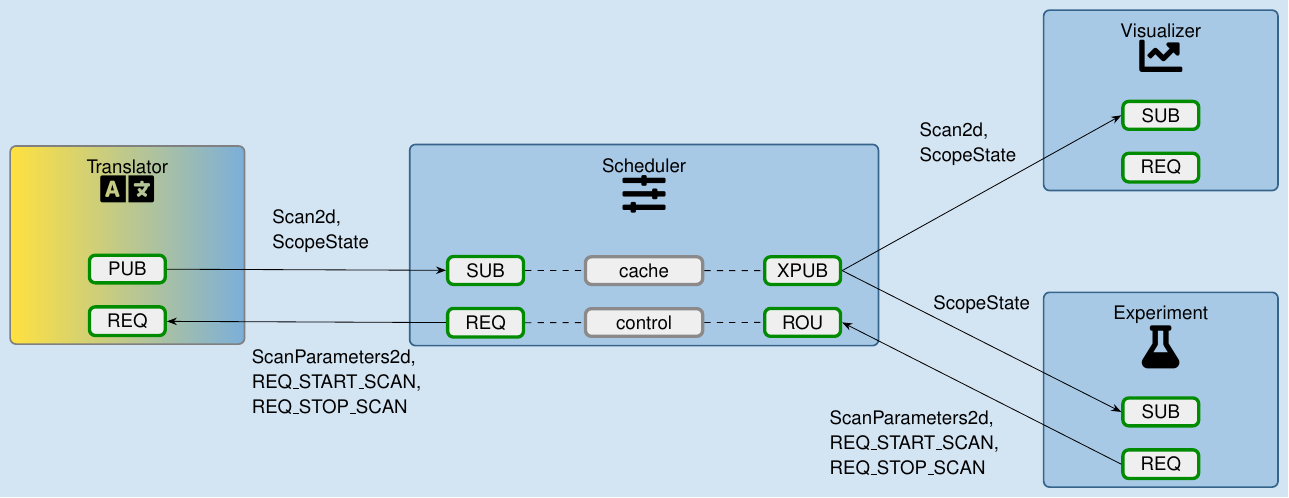}
  }
  \caption{There are two main communication paths in \texttt{afspm}: a publisher-subscriber one, and a request-control one.
    The publisher-subscriber path allows the microscope to send information about its state, which all components can subscribe to.
    The control-request path allows components to take control of the microscope and request it perform actions or change its configuration.
    The Microscope Scheduler caches data in the publisher-subscriber path (to send out to new components), and mediates control of the microscope in the control-request path.}
  \label{fig:afspm_control}
\end{figure}

Communicating between the SPM and other components is decoupled in two paths: a publisher-subscriber path to receive events from the SPM, and a control path to handle requests to the SPM (see Figure \ref{fig:afspm_control}) .
Unless a component exists only to handle `intermediary' data (between components), it will likely subscribe to at least one message `topic' from -- and send at least one type of control request to -- the SPM.
In this example diagram, a \texttt{Scan Decider} component is subscribed only to \texttt{ScopeState} messages from the SPM, and sends requests to start a scan, stop a scan, and change the \texttt{ScanParameters2d}.
This example component concerns itself only with deciding where to scan, and has some internal logic to determine these locations.
A \texttt{Visualizer} component is subscribed only to \texttt{Scan2d} and \texttt{ScopeState} messages from the SPM, and does not send control requests.
As we may expect, this component simply visualizes the latest scans to the user.

On the publisher-subscriber path, the SPM publishes different message types according to what changes have occurred.
For example, if the scanning state of the SPM has changed, a new \texttt{ScopeState} message is sent; if a scan has finished, the new scan is read from disk into a \texttt{Scan2d} message and published.
We have predefined most expected standard message types, but any user may define their own message type and compile it using the protobuffers utility.
To better understand the publisher-subscriber data structure schemas used and overall message format, review the sample schemas in Figure \ref{fig:pub_sub_schemas_format}.
Components need not receive all of the myriad message types the SPM may send --- they may subscribe only to the particular message types that interest them.
These message types are differentiated in \texttt{ZeroMQ} via simple string keys denominated as `envelopes', and the mapping from a given message type to a given envelope is configurable in \texttt{afspm} (although a reasonable default is provided).
The user may define what envelopes a given component connects to during configuration of an experiment, simplifying handling of new messages.

\begin{figure}
  \centering
  \begin{subfigure}[b]{0.49\textwidth}
    \inputminted[firstline=1,lastline=15]{protobuf}{./minted_src/scan.proto}
    \caption{2D Scan Parameters Schema}
    \label{code:scan_parameters}
    \bigskip
  \end{subfigure}
  \hfill
  \begin{subfigure}[b]{0.49\textwidth}
    \inputminted[firstline=17,lastline=23]{protobuf}{./minted_src/scan.proto}
    \caption{2D Scan Schema}
    \label{code:scan}
    \bigskip
  \end{subfigure}
  \begin{subfigure}[b]{0.9\textwidth}
    \begin{bytefield}{50}
      \bitbox[]{10}{Frame 1} & \bitbox{20}{\texttt{`Scan2d'}} & \bitbox[]{20}{\textit{Envelope}} \\
      \bitbox[]{10}{Frame 2} & \bitbox{20}{\texttt{Scan2d Data}} & \bitbox[]{20}{\textit{Serialized Data Structure}}
    \end{bytefield}
    \caption{Publisher-Subscriber Message Format}
    \label{diagram:published_message}
  \end{subfigure}
  \caption{Sample Publisher-Subscriber Schemas and Message Format --- (a) spatial and data aspects both contain units defined for proper interpretation and manipulation (see Figure \ref{code:geometry} for some of these geometric primitives).
  (b) the \texttt{params} and \texttt{channel} attributes indicate the data collected, the raw data is stored in \texttt{values}, and \texttt{timestamp} and \texttt{filename} distinguish it in time and on disk.
  (c) messages published from the SPM are prepended an envelope that may be used to filter message types of interest.
  The default envelope creator uses the message name as the envelope string.}
  \label{fig:pub_sub_schemas_format}
\end{figure}

In the control path, components send requests to the SPM and receive responses accordingly.
The Microscope Scheduler functions as a router between the Microscope Translator and all other components, ensuring that only one component is controling the SPM at a given instance in time.
Assuming a component is in control, it may send any of the defined control requests to the SPM; the SPM (or Translator) must respond to every request with an appropriate control response.
A component not in control may request control and add/remove `flags' indicating a detected problem with the experiment (for example, a tip characterization issue).
By this mechanism, a component may pause regular running of the experiment: if one or more problems are flagged, the Scheduler removes the current component in control.
At this stage, only components indicating they resolve one of the currently logged problems may obtain control.

When running an experiment, it is likely the automated components will at one time or another behave against one's expectations.
To account for this (and generally give the user control over the experiment), we define a \texttt{ControlMode} under which the experiment runs.
This mode can be either \texttt{AUTOMATED} or \texttt{MANUAL}, with the former being the default.
At any time, the user may switch to \texttt{MANUAL} mode, which disengages any \texttt{afspm} components in control of the microscope and refuses any requests to gain control.
When the user is satisfied and willing to resume automation, they may switch back to \texttt{AUTOMATED} mode, allowing components to reconnect and continue the experiment.

\begin{figure}
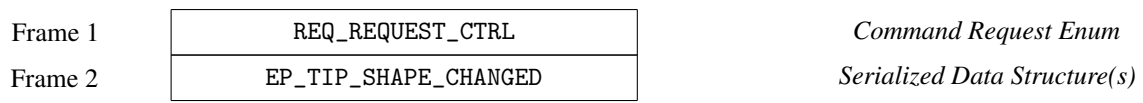

  \centering
  \begin{subfigure}[b]{0.4\textwidth}
    \inputminted[firstline=1,lastline=18]{protobuf}{./minted_src/control.proto}
    \caption{Control Request Schema}
    \label{code:control_request}
    \bigskip
  \end{subfigure}
  \hspace{7em}
  \begin{subfigure}[b]{0.4\textwidth}
    \inputminted[firstline=20,lastline=34]{protobuf}{./minted_src/control.proto}
    \caption{Control Response Schema}
    \label{code:control_response}
    \bigskip
  \end{subfigure}
  \begin{subfigure}[b]{0.4\textwidth}
    \inputminted[firstline=42,lastline=48]{protobuf}{./minted_src/control.proto}
    \caption{Experiment Problem Schema}
    \label{code:experiment_problem}
    \bigskip
  \end{subfigure}
  \hspace{7em}
  \begin{subfigure}[b]{0.4\textwidth}
    \inputminted[firstline=36,lastline=40]{protobuf}{./minted_src/control.proto}
    \caption{Control Mode Schema}
    \label{code:control_mode}
    \bigskip
  \end{subfigure}
  \begin{subfigure}[b]{0.99\textwidth}
    \begin{bytefield}{60}
      \bitbox[]{10}{Frame 1} & \bitbox{20}{\texttt{REQ\_REQUEST\_CTRL}} & \bitbox[]{30}{\textit{Command Request Enum}} \\
      \bitbox[]{10}{Frame 2} & \bitbox{20}{\texttt{EP\_TIP\_SHAPE\_CHANGED}} & \bitbox[]{30}{\textit{Serialized Data Structure(s)}}
    \end{bytefield}
    \caption{Control Request Message Format}
    \label{diagram:control_request_example}
  \end{subfigure}
  \caption{Sample Control Schemas and Message Format --- (a) and (b) indicate schemas for possible control requests and response.
  (c) shows various experiment problems that may be flagged.
  (d) indicates the mode the experiment may be in, with \texttt{AUTOMATED} being the default (where automation is running).
  (e) shows a sample control request message, which consists of the control request enum and potentially one or more additional data structures, dependent on the request.
  In this case, a \texttt{REQ\_REQUEST\_CTRL} request is being sent, which demands an \texttt{ExperimentProblem} enum indicating the problem this requestor solves (if any).
  In this particular case, it is \texttt{EP\_TIP\_SHAPE\_CHANGED}.}
\label{fig:control_schemas_format}
\end{figure}

For an overview of control path message types and the message format, see Figure \ref{fig:control_schemas_format}.

\subsection{Caching and the Microscope Scheduler}

The mapping from message type to envelope is also critical in the Microscope Scheduler logic: as it sits between the translator and other components, the scheduler receives SPM events from the Microscope Translator and republishes them.
In doing so, it holds the data in a publisher-subscription `cache' mechanism, which both caches data as necessary and converts it from an incoming `envelope' to an outgoing `envelope'.
For caching data, data is stored into a <key:value> mapping according to user-defined logic.
When a new component connects to the scheduler's publishing socket, all of the messages in its cache are sent out.
The goal is to hold as much history (per message type) as needed for all components to function properly regardless of when they were created.
While the SPM should have a simple one-to-one mapping between message type and envelope, we may desire more complex logic when sending these messages to other components in the experiment -- this is what is definable for the scheduler.
As a simple example, one experiment may involve scanning a surface at two different `granularities': a wide scan area resolution, to visualize the larger region, and a narrow-scan area resolution, to focus on particular adatoms of interest.
In this case, we would be interested in caching both wide-area and narrow-area scans differently, so they both may be sent to our components when they create.
As the message type of these is the same (i.e. \texttt{Scan2d}), we need a more involved message type to envelope mapping.

\subsection{Handling Units}

The various SPMs will likely send data defined with different units and developed automation components may desire to calculate in particular units.
Handling these various units is thus important for reliable functioning of any experiment.
It is worth repeating that all data-based schema have defined units, and afspm provides helper methods to allow easy converting from received to desired units.
We argue that each defined component should output data using its internally desired units, and convert any received inputs into these same units.
The responsibility for conversion thus lies with each individual component.
Helper methods are provided to perform these conversions.

\subsection{Spawning and Monitoring Components}

The components to be run during an experiment can be defined in a single configuration file, and `spawned' (or created) all-at-once or piecemeal via the command line.
The term spawn here explicits that each component is created as a child process, rather than a thread within a single process.
In doing so, we ensure that faulty code in one component does not crash all components of the experiment, as each process has its own memory.
Additionally, this allows for flexibility in concurrent computing: if the components are configured to communicate over TCP/IP sockets, they can be spawned on multiple computing devices.

The parent process which spawns components additionally monitors them, to ensure no frozen or crashed component is blocking the experiment.
It accomplishes this by monitoring for `heartbeats', explicit messages passed at a defined frequency by each component to the parent process.
If a component has not sent a heartbeat within a defined timeframe, the monitor assumes it has frozen or crashed.
It proceeds to force kill the process and respawn it.
This highlights the importance of a properly defined cache: upon respawn, a component must receive at least as much recent information as it needs to produce an output.
Otherwise, the experiment will continue to be blocked.

\subsection{Experiment Configuration}

Our framework uses a TOML-based configuration file \citep{tomlTOMLTomaposs} to define components of an experiment, their settings, and the manners by which they communicate.
Its general structure can be best understood from a sample configuration, shown in Figure \ref{fig:configuration_file}.
The file can be decoupled into three sections: general variables, intermediary classes, and components.
General variables are the network sockets by which components communicate, as well as any additional variables that will be used to configure intermediary classes or components.
Note the structure of a \texttt{ZeroMQ} socket contains the `network transport' (e.g. `tcp' for TCP/IP or `ipc' for inter-process communication) followed by the differentiable id for the socket itself (in this case, a TCP/IP address and socket). Intermediary classes can be thought of as `sub-components', classes used to set up the communication paths.
In this example, they configure the control and publisher-subscriber paths between all the components.
The components themselves are defined by having a boolean \texttt{component} attribute, with the \texttt{class} attribute used to determine how to load Python-based components.
Note that string values are cross-referenced with keys to perform variable replacement (e.g. \texttt{`pub\_url'} is replaced by \texttt{`tcp://127.0.0.1:9000'} in \texttt{translator\_pub}).

\begin{figure}
  \centering
  \begin{subfigure}[b]{0.42\textwidth}
    \inputminted[firstline=1,lastline=16]{toml}{./minted_src/config.toml}
    \caption{General Variables}
    \label{code:config_general_variables}
    \bigskip
  \end{subfigure}
  \hfill
  \begin{subfigure}[b]{0.53\textwidth}
    \inputminted[firstline=18,lastline=33]{toml}{./minted_src/config.toml}
    \caption{Intermediary Classes}
    \label{code:config_intermediary_classes}
    \bigskip
  \end{subfigure}
  \begin{subfigure}[b]{0.95\textwidth}
    \inputminted[firstline=35,lastline=58]{toml}{./minted_src/config.toml}
    \caption{Components}
    \label{code:config_components}
  \end{subfigure}
  \caption{Excerpts from a sample configuration file, which can be divided into general variables defined at the beginning (a); intermediary classes (primarily input/output communication classes) (b); and components with their configuration (c).}
\label{fig:configuration_file}
\end{figure}

\begin{figure}
  \centering
  \begin{subfigure}[b]{0.35\textwidth}
    \inputminted[firstline=1,lastline=9]{proto}{./minted_src/geometry.proto}
  \end{subfigure}
  \hspace{1em}
  \begin{subfigure}[b]{0.35\textwidth}
    \inputminted[firstline=11,lastline=20]{proto}{./minted_src/geometry.proto}
  \end{subfigure}
    \caption{Sampling of geometric schemas used by our SPM schemas.}
    \label{code:geometry}
\end{figure}

\newpage
\section{Drift Correction}

\subsection{Drift Estimation}

\begin{figure}
  \centering
  \resizebox{\textwidth}{!}{  
    \includegraphics[width=\textwidth]{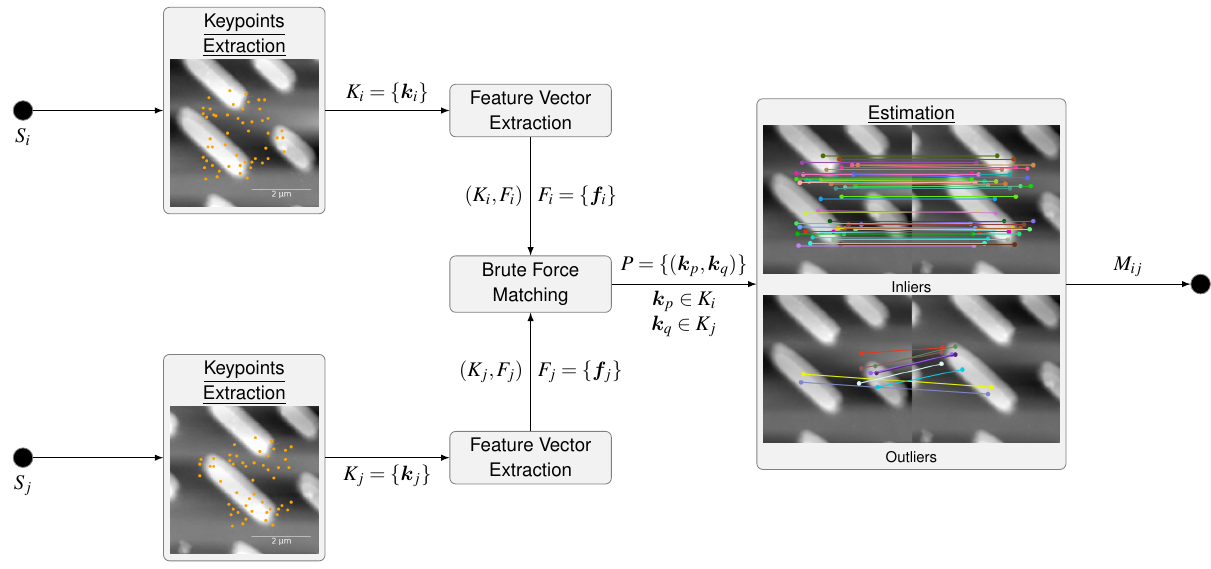}
  }
  \caption{In image registration, the linear mapping necessary to convert an image $S_{j}$ to the same coordinate system as an image $S_{i}$ is estimated.
    This is achieved by estimating 2D key points $\bm{k}_{i}$ and $\bm{k}_{j}$ in the two images (shown as orange circles); extracting associated feature vectors $\bm{f}_{i}$ and $\bm{f}_{j}$; performing brute force matching to pairs of keypoints $(\bm{k}_{p}, \bm{k}_{q})$ that match each other in both images; and estimating the mapping $M_{ij}$ based on these keypoint pairs via an outlier-robust method.
    In the diagram, the paired keypoints have matching colours and lines connecting them.
    Inlier and outlier pairs are shown separately.  }
  \label{fig:drift_estimation}
\end{figure}

Various mechanisms have been developed for estimating drift: tracking common features between two images (as done in \citep{diao2023automatic}, and the aperiodic method in \citep{dickbreder2023undrift}), performing a cross-correlation between two images (such as the periodic approach in \citep{dickbreder2023undrift}), and active feedback approach involving dithering above an atom \cite{swartzentruber1996direct}.
Our approach to estimate drift offsets between a pair of scans uses a features-based approach and follows standard image registration techniques for transforming images into a common coordinate system.
The goal is to estimate a linear mapping that transforms a scan $S_{j}$ to the same coordinate system as a scan $S_{i}$, where the form of the mapping is constrained by the process by which the transformation occured; this could vary from a simple 2D translation up to a more complicated transformation such as affine ones.
The general algorithmic flow of this process can be seen in Figure \ref{fig:drift_estimation}, where two such example scans $S_{i}$ and $S_{j}$ can be seen on the far left.
Throughout this explanation, we will highlight the individual choices we have made in our implementation, and compare them to that of \cite{diao2023automatic}, which uses a similar image registration technique.

The mapping is estimated by first determining salient 2D points in each image, creating two sets $K_{i} = \{ \bm{k}_{i} \vert \bm{k}_{i} \in \mathbb{R}^{2} \}$ and $K_{j} = \{ \bm{k}_{j} \vert \bm{k}_{j} \in \mathbb{R}^{2} \}$, such that $\bm{k_{i}}$ and $\bm{k_{j}}$ are 2D points.
We also note that $\mathrm{size}\left(K_{i}\right) = n$ and $\mathrm{size}\left(K_{j}\right) = m$, where $\mathrm{size}\left(\cdot\right)$ is a function that returns the size of the set -- meaning they have $n$ and $m$ keypoints, respectively.
Two such sets of keypoints are shown in orange atop the grayscale visualizations of the scans, in the `Keypoints Extraction' blocks on the left of Figure \ref{fig:drift_estimation}.
The extraction of key points attempts to find salient points in the image that adhere to certain conditions.
For example, it is common to extract features from the image at various length scales, considering only key points that appear at the various scales; this imbues some degree of scale-invariance to the points.
For each of these keypoints, we then extract a feature vector which we expect describes the key point reliably, resulting in two sets $F_{i} = \{ \bm{f}_{i} \vert \bm{f}_{i} \in \mathbb{R}^{o} \}$ and $F_{j} = \{ \bm{f}_{j} \vert \bm{f}_{j} \in \mathbb{R}^{o} \}$ with feature vectors $\bm{f}_{i}$, $\bm{f_{j}}$ of size $o$, where $\mathrm{size}\left(F_{i}\right) = n$, $\mathrm{size}\left(F_{j}\right) = m$.
With two sets of feature vectors corresponding to the two images we then perform brute force matching, where feature vectors in $F_{i}$ are matched to those in $F_{j}$ for which the vector is closest.
This results in keypoint pairs $P = \{ (\bm{k}_{p}, \bm{k}_{q}) \vert \bm{k}_{p} \in K_{i}, \bm{k}_{q} \in K_{j}\}$, where $\mathrm{size}\left(P\right) \leq n, \mathrm{size}\left(P\right) \leq m$.
Further consistency can be achieved by cross-checking, i.e. ensuring that each of the feature vectors in a pair is the best match of the other.
In our approach, we used brute force matching with cross-checking.
This matching process is showing in the center of Figure \ref{fig:drift_estimation}.

For the detection of keypoints and extraction of feature vectors a number of different methods exist, among them SIFT \citep{lowe1999object}, SURF \citep{bay2006surf}, BRIEF \citep{calonder2010brief}, ORB \citep{rublee2011orb}, KAZE \citep{alcantarilla2012kaze}, and AKAZE \citep{alcantarilla2011fast}.
These methods differ in the feature vectors they extract and the mechanism by which they attempt to guarantee reproducible keypoints following various transformation types. 
In our implementation, we integrated SIFT, BRIEF, and ORB, selecting BRIEF after some minimal evaluation due to the high number of detected keypoints minimizing the risk of a mapping not being estimated.
We acknowledge that such a heuristic is simplistic, as the detected keypoints could be less reproducible and thus lead to poor mapping estimates.
In \cite{diao2023automatic}, AKAZE estimation with KAZE features were used, which provides higher localization accuracy (i.e. keypoints will remain in the correct position at various scales) at a higher computational cost.

Once we have a list of keypoint pairs, we can move on to the estimation of drift offset.
A simple mechanism to do so would be to find the least-squares offset of the data.
This risks introducing a bias associated with outlier keypoint pairs (due, for example, to improper matching).
To attempt to minimize this, one can use one of a number of mechanisms to attempt to detect and eliminate outliers from the estimation process.
In our implementation, we use the RAndom SAmpling Consensus algorithm (RANSAC) \citep{derpanis2010overview} on top of a standard least-squares fitting.
RANSAC attempts to minimize the influence of outliers by iteratively: (a) subsampling from the set of data points; (b) performing a model fit with this subset; (c) determining inliers and outliers to the fit model based on a loss function (such as the squared error between data points and model); and (d) refitting the model using the inlier set (also termed the consensus set).
After each iteration, RANSAC will keep either the current model fit or the prior fit, whichever is better.
It will end early if the consensus set is a sufficiently large proportion of the total dataset.
Otherwise, it will stop after a preset number of iterations.
In the end, the mapping outputted by RANSAC is used as our estimated mapping.
An example of keypoint pairs and their computed classification (inlier or outlier) computed using this approach can be seen in the `Estimation' block to the right of Figure \ref{fig:drift_estimation}.
Here, the two scans are visualized side by side, with keypoint pairs drawn with the same colour, and the lines connecting the keypoint pairs indicating the translation offset between them.
In the Figure, we can see both the matched inlier keypoint pairs in the top image, and the outlier ones in the bottom image.

In RANSAC, there are two key parameters that must be optimized for the problem at hand: the loss function and threshold value chosen to define outliers relative to an estimate, and the number of iterations permitted before automatic stopping.
For the loss function, we use the default squared error between data points, with a residual threshold defined as a percentage of the scan resolution.
In doing so, we aim to define thresholds that are invariant to image resolution, with our default threshold being set at 5\% after some minimal evaluation.
We did not vary the number of iterations, which defaulted to 100 with our backend.
We acknowledge that further experimentation would be beneficial to optimize the system for a given experiment (for example, considering the inlier/outlier ratio as a function of these parameters).
However, our focus was on validating the usefulness of our framework and the process of writing generic components.

In \cite{diao2023automatic}, outliers are eliminated by a k-means clustering approach, where the data are separated into $N$ clusters, with $N$ set so as to ensure that all clusters have data points within a squared error threshold from the cluster centroid.
Thus, we can see that both our implementation and that of \cite{diao2023automatic} use a distance-based threshold to determine outliers, with the value of this threshold being decided via some heuristic (in our case, as a percentage of the scan resolution).

\subsection{Drift Correction Algorithm}

Our approach to correcting for drift considers estimated drift mappings $M_{T \rightarrow S, ij}$ from the TCS to the SCS between two scans $S_{i}$ and $S_{j}$, respectively, by identifying key features in each image.
We use these to estimate a total mapping $M_{T \rightarrow S, tot}$, where the subscript $tot$ indicates it is the total correction at an instance in time relative to the start of the experiment.
For simplicity, we denominate $M_{ij} := M_{T \rightarrow S, ij}$ and
$M_{tot} := M_{T \rightarrow S, tot}$.
We are functioning in 2D homogeneous coordinates $\bm{x} = \left[ x_{0}, x_{1}, 1 \right]^{\intercal}$ and with mappings of the form $M = \left[ R_{\mathrm{3\times2}} \vert \bm{t}_{\mathrm{3\times1}} \right]$ containing both the $\mathrm{3\times2}$ rotation matrix $R$ and the $\mathrm{3\times1}$ translation vector $\bm{t}$.
Note that at a given instance in time:
\begin{equation}
  \bm{x}_{S} = M_{tot} \cdot \bm{x}_{T}
\end{equation}
where $\bm{x}_{S}$ represents a point in the SCS and $\bm{x}_{T}$ the same point in the TCS, and $M_{tot}$ varies as a function of time due to the varying thermal drift.
To correct, we assume $M_{tot} = \left[ I_{\mathrm{3\times2}} \vert \bm{0}_{\mathrm{3\times1}}\right]$ at the start of an experiment, with $I_{\mathrm{3\times2}}$ the identity matrix and $\bm{0}_{\mathrm{3\times1}}$ a zero vector.
In other words, we assume that the SCS and TCS are equivalent at the start of the experiment, with our correction mapping correcting the drift we detect and track over said experiment via our estimated mappings $M_{ij}$ between scans at every scan.
Assuming we are attempting to update $M_{tot}$ at every scan, we can imagine that $M_{ij}$ indicates the drift at a time step $t_{k}$, such that $\Delta M_{tot, k} = M_{ij}$,where the value is positive because it is a mapping from our later scan $S_{j}$ to our earlier scan $S_{i}$ and thus corrects for the drift that occurred.
In other words, in this scheme the drift estimated between two scans represents the change in the mapping over that time duration.
The mapping update is then:
\begin{equation}
  M_{tot,k} = M_{tot,k-1} \cdot \Delta M_{tot, k}
\end{equation}
where we are chaining coordinate system transformations: $\Delta M_{tot, k}$ is a mapping from the coordinate system at time $k$ to that at $k-1$, and $M_{tot,k-1}$ is the complete mapping from the coordinate system at time $k-1$ to that at the start of the experiment.
Such a correction may accurately update the mapping between scans, but does not predict future drift.
If the drift rate is sufficiently large relative to the time between scans, there may be insufficient common surface between scans to estimate a change in the drift mapping.
Because of this, we additionally perform a first-order prediction at each time step:
\begin{equation}
  \widehat{\Delta M}_{tot, k} = \left( \frac{\Delta M_{tot, k-1}}{\Delta t_{(k-2, k-1)}} \right) \cdot \Delta t_{(k-1, k)} = \dot{M}_{tot,k-1} \cdot \Delta t_{(k-1, k)}
  \label{eq:mapping_delta_prediction}
\end{equation}
where $\Delta t_{(k-1, k)}$ is the time between scans $S_{k-1}$ and $S_{k}$.
The update step in this scheme is then:
\begin{equation}
  M_{tot,k}' = M_{tot,k-1} \cdot \widehat{\Delta M}_{tot, k}
  \label{eq:mapping_predictive_update}
\end{equation}
\begin{equation}
\begin{split}
  M_{tot,k} = M_{tot,k}' \cdot \widetilde{\Delta M}_{tot, k} &= M_{tot,k-1} \cdot \widehat{\Delta M}_{tot, k} \cdot \widetilde{\Delta M}_{tot, k} \\
  &= M_{tot,k-1} \Delta M_{tot, k}
\end{split}
  \label{eq:mapping_estimation_update}
\end{equation}
where $M_{tot,k}'$ is used to determine the scan position of $S_{k}$ at the start of the scan, and $M_{tot,k}$ is the updated mapping considering the drift we detected after the scan.
Here, $\Delta M_{tot, k} = \widehat{\Delta M}_{tot, k} \cdot \widetilde{\Delta M}_{tot, k}$, i.e. $\widetilde{\Delta M}_{tot, k}$ represents the delta between the true drift and the first-order prediction of drift for scan $S_{k}$.
Since we are using our prediction $\widehat{\Delta M}_{tot,k}$ when starting the scan in equation \eqref{eq:mapping_predictive_update}, $\widetilde{\Delta M}_{tot, k}$ acts as a proxy for $M_{ij}$ in this scheme.

\subsubsection{Drift Correction Example}
\begin{figure}
  \centering
  \begin{subfigure}[b]{\textwidth}
    \resizebox{\textwidth}{!}{
      \includegraphics[width=\textwidth]{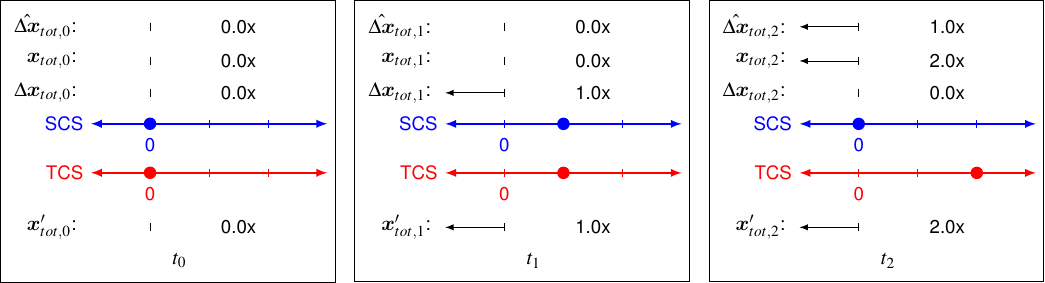}
    }
    \caption{}
    \label{fig:drift_correction_mechanism_start}
  \end{subfigure}
  \hfill
  \begin{subfigure}[b]{\textwidth}
    \resizebox{\textwidth}{!}{
      \includegraphics[width=\textwidth]{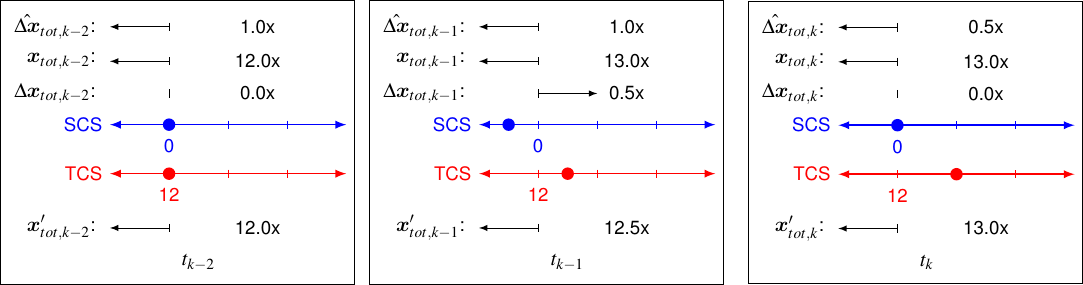}
    }
    \caption{}
    \label{fig:drift_correction_mechanism_rate_change}
  \end{subfigure}
  \caption{Illustration of drift correction mechanism with a one-dimensional coordinate system. \ref{fig:drift_correction_mechanism_start} shows drift correction estimated on startup, where we begin with an assumed null vector for our drift. \ref{fig:drift_correction_mechanism_rate_change} shows an updating of our predictive step due to a change in drift rate.}
  \label{fig:drift_correction_mechanism}
\end{figure}

The correction mechanism described above is exemplified in Figure \ref{fig:drift_correction_mechanism}, where a one-dimensional coordinate system is used resulting in positions $x$ rather than mappings $M$.
For both \ref{fig:drift_correction_mechanism_start} and \ref{fig:drift_correction_mechanism_rate_change}, the system is continuously scanning a region with a top-left position of the origin in the scanning coordinate system (SCS) while a drift occurs.
In \ref{fig:drift_correction_mechanism_start}, the mechanism is shown on startup.
We begin with a $x_{tot}$  of zero;
since we have not detected any prior drift, our prediction $\hat{\Delta x}_{tot,0}$ at $t_{0}$ is also zero.
As we do not have a pair of scans with which to estimate drift, $\Delta x_{tot,0}$ is zero and our updated mapping $x_{tot,0}'$ remains zero.
At $t_{1}$, $\hat{\Delta x_{tot,1}}$ and $x_{tot,1}$ remain zero due to no predicted drift rate.
However, we now obtain a scan pair which we can use for drift estimation.
Our drift estimation between $S_{1}$ and $S_{2}$ shows a value for $\Delta x_{tot, 1}$ of $(1.0)$, implying the position of $S_{2}$ in the SCS is $(1.0)$.
In reality, this is indicative of drift, which we correct for with an update of $x_{tot,0}'$ to $(1.0)$.
For $t_{2}$, we first compute our expected TCS coordinates for our SCS position of $(0.0)$ using the predictive estimate, resulting in a value of $x_{tot,2}$ of $(2.0)$ which we send to the microscope.
Upon drift estimation, we detect no drift as $\Delta x_{tot,2}$, implying that our drift rate is constant and our predicted drift was correct (in this example, there is no noise).
Our updated mapping thus matches our initial predictive estimate.
In \ref{fig:drift_correction_mechanism_rate_change}, we witness the drift estimation mechanism adapting to a change in drift rate.
At $t_{n}$, the drift rate has been correctly estimated and is being tracked accurately (as in $t_{2}$ from \ref{fig:drift_correction_mechanism_start}).
At $t_{n-1}$, the actual drift rate has decreased to $(0.5)$ units per time step.
Our drift estimation thus detects a value for $\Delta x_{tot,n-1}$ of $(-0.5)$, resulting in our total correction being updated to $(12.5)$ (from $(13.0)$) via $x_{tot,n-1}'$.
At $t_{n-2}$, our predicted mapping matches reality, resulting in a $\Delta x_{tot,n-2}$ of zero (as in $t_{2}$ in \ref{fig:drift_correction_mechanism_start}).

\newpage
\section{Additional Results}

\subsection{Drift Correction}

\begin{figure}
  \centering
  \includegraphics[width=\textwidth]{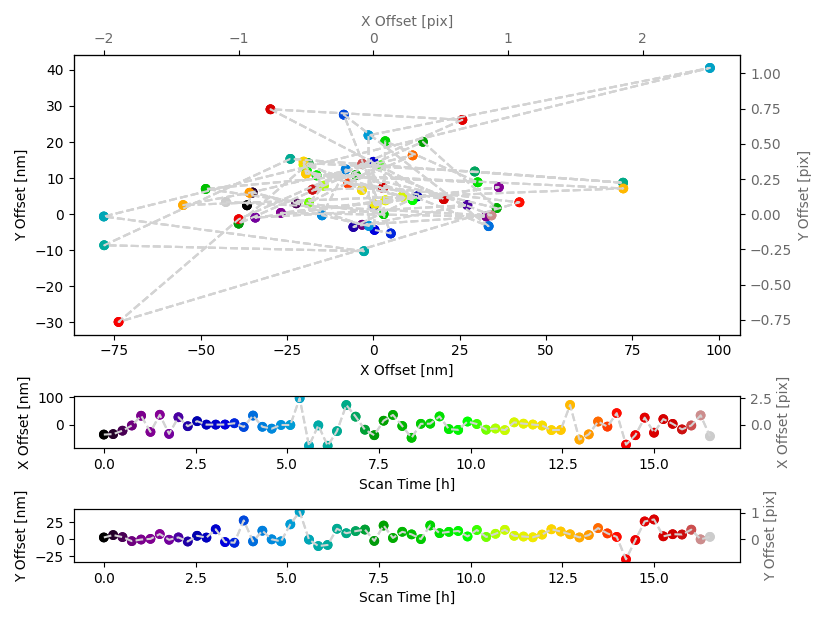}
  \caption{Visualization of drift offset $\bm{x}_{tot}$ run offline, on the scans with drift correction enabled.}
  \label{fig:drift_correction_results_offline}
\end{figure}

\begin{figure}
  \centering
  \includegraphics[width=\textwidth]{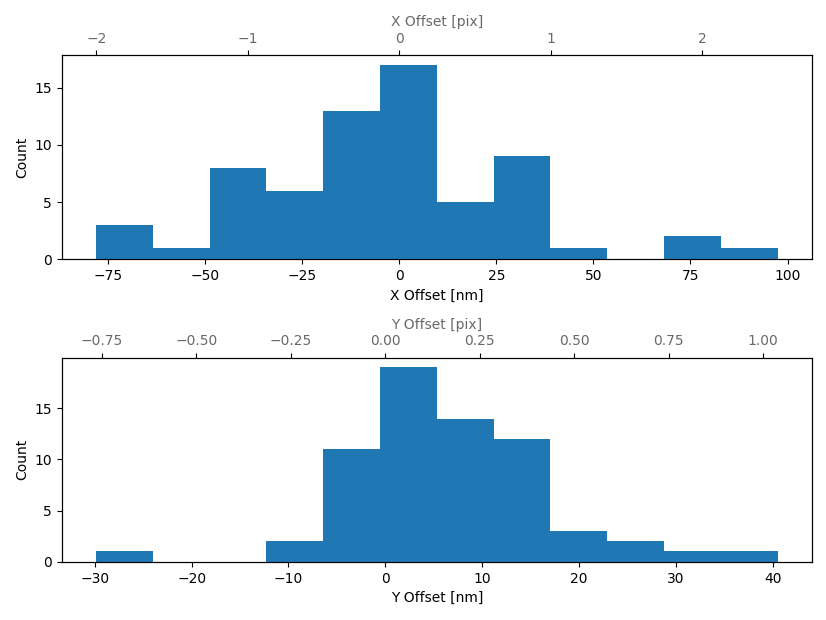}
  \caption{Histogram of the drift offset $\bm{x}_{tot}$ run offline, on the scans with drift correction enabled.
  The algorithm used to set the number of bins was the Friedman-Diaconis estimator \citep{freedman1981histogram}.}
  \label{fig:drift_correction_results_offline_hist}
\end{figure}

\begin{figure}
  \centering
  \includegraphics[width=\textwidth]{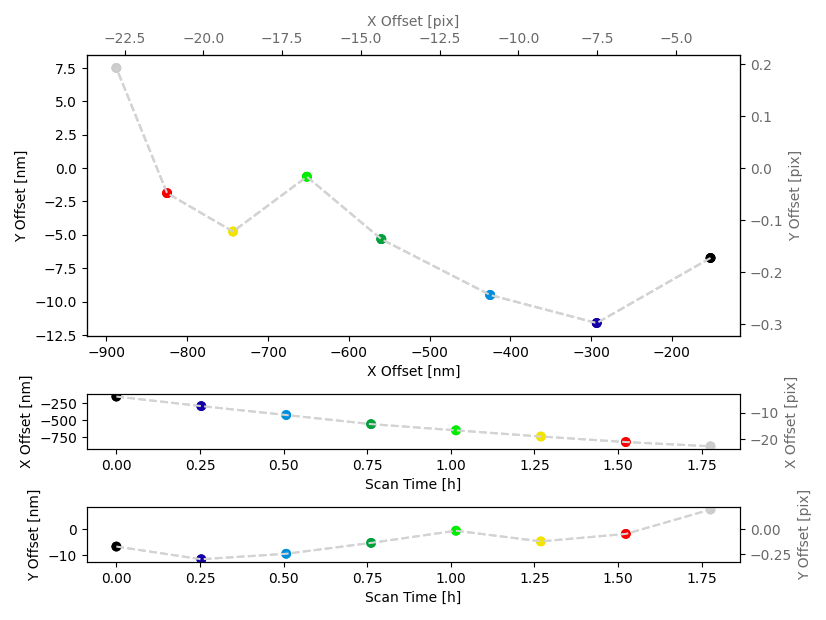}
  \caption{Visualization of drift offset $\bm{x}_{tot}$ run offline, on the scans with drift correction disabled.}
  \label{fig:passive_drift_results_offline}
\end{figure}

We ran an offline version of the correction, where we output the offset $\bm{x}_{tot}$, which is visible in Figure \ref{fig:drift_correction_results_offline}.
In this case, we are only measuring the absolute offset between the first scan and each subsequent one; if there were zero noise in our correction, we would expect $\bm{x}_{tot} = \bm{0}$ throughout the experiment.
As we see, the drift appears to have been corrected throughout the experiment.
While there are some jumps in estimated drift (attributable to errors in our drift prediction, where the drift changed), generally the drift visualized here is very low, attributable to minor errors in our linear prediction and error in our estimation.
This is best visualized when plotting these errors as histograms, as seen in Figure \ref{fig:drift_correction_results_offline_hist}.
In Figure \ref{fig:passive_drift_results_offline}, we see a similar offline estimation for the collection where drift correction was disabled.
This gives us a view of the general drift we are to expect -- though we caution this collection was of a short duration and there is significant variation in the drift we saw over time.
Nonetheless, we see the drift is comparable to the drift our estimator detected.

\end{appendix}
\end{document}